\documentclass[twoside,english,18pt]{elsarticle}
\usepackage[T1]{fontenc}
\usepackage[latin9]{inputenc}
\usepackage{geometry}
\usepackage{color}
\usepackage{float}
\usepackage{amsmath}
\usepackage{amsthm}
\usepackage{amssymb}
\usepackage{graphicx}
\usepackage{setspace}
\makeatletter
\theoremstyle{definition}
\newtheorem{defn}{\protect\definitionname}
\theoremstyle{plain}
\newtheorem{lem}{\protect\lemmaname}
\theoremstyle{remark}
\newtheorem{rem}{\protect\remarkname}
\theoremstyle{plain}
\newtheorem{thm}{\protect\theoremname}
\ifx\proof\undefined\
  \newenvironment{proof}[1][\proofname]{\par
    \normalfont\topsep6\p@\@plus6\p@\relax
    \trivlist
    \itemindent\parindent
    \item[\hskip\labelsep
          \scshape
      #1]\ignorespaces
  }{%
    \endtrivlist\@endpefalse
  }
  \providecommand{\proofname}{Proof}
\fi

\usepackage[caption=false,font=footnotesize]{subfig}

\usepackage{amstext}
\usepackage{pdfsync}
\usepackage{amsthm}

\usepackage{epstopdf}

\usepackage{tikz}
\usetikzlibrary{arrows}
\usetikzlibrary{shadows}
\tikzset{
  every overlay node/.style={
    draw=white,anchor=north west,
  },
}
\usetikzlibrary{positioning, decorations.pathmorphing}

\usepackage{hyperref}
\@ifundefined{rangeHsb}{\usepackage{xcolor}}{}

\usepackage{algorithm,algpseudocode}

\usepackage{lipsum}

\usepackage{scrlayer-scrpage}
\clearpairofpagestyles
\makeatother

\usepackage{babel}
\providecommand{\definitionname}{Definition}
\providecommand{\lemmaname}{Lemma}
\providecommand{\remarkname}{Remark}
\providecommand{\theoremname}{Theorem}

\begin{document}

\begin{frontmatter}{}

\title{Dynamic Event-Triggered Consensus of Multi-agent System on Matrix-weighted
Networks}

\author[sjtu1,sjtu2,sjtu3]{Lulu~Pan }

\author[sjtu1,sjtu2,sjtu3]{Haibin~Shao\corref{cor1} }

\author[sjtu1,sjtu2,sjtu3]{Dewei Li }

\author[sdu]{Lin Liu }

\fntext[fn2]{This work is supported by the National Natural Science Foundation
of China (Grant No. 62103278, 61973214, 61963030) and Natural Science
Foundation of Shanghai (Grant No. 19ZR1476200) and in part by Scientific
Research Funding of Shanghai Dianji University (Grant No. B1-0288-21-007-01-023)
}

\cortext[cor1]{Corresponding author.}

\address[sjtu1]{Department of Automation, Shanghai Jiao Tong University, Shanghai,
200240, China}

\address[sjtu2]{Key Laboratory of System Control and Information Processing, Ministry
of Education of China, Shanghai 200240, China }

\address[sjtu3]{Shanghai Engineering Research Center of Intelligent Control and Management,
Shanghai, 200240, China}

\address[sdu]{School of Electronic Information Engineering, Shanghai Dianji University,
Shanghai 201306, China}
\begin{abstract}
This paper examines the event-triggered consensus of the multi-agent
system on matrix-weighted networks, where the interdependencies among
higher-dimensional states of neighboring agents are characterized
by matrix-weighted edges in the network. Specifically, a novel distributed
dynamic event-triggered coordination strategy is proposed for this
category of generalized networks, in which an auxiliary system is
employed for each agent to dynamically adjust the triggering threshold,
which plays an essential role in guaranteeing that the triggering
time sequence does not exhibit Zeno behavior. Distributed event-triggered
control protocols are proposed to guarantee leaderless and leader-follower
consensus for multi-agent systems on matrix-weighted networks, respectively.
Remarkably, the spectrum of matrix-valued weights is crucial in event-triggered
mechanism design for matrix-weighted networks, generalizing those
results only applicable for scalar-weighted networks. The proposed
approach allows each agent to broadcast and receive information only
at its triggering instants. Finally, simulation examples are provided
to demonstrate the theoretical results.
\end{abstract}

\end{frontmatter}{}

\section{Introduction}

Reaching a consensus is a paramount routine in distributed coordination
of multi-agent systems \citet{mesbahi2010graph,olfati2007consensus,degroot1974reaching}.
Although the consensus problem has been extensively investigated,
the ties among agents are assumed to be characterized by scalar-weighted
networks, which fail in characterizing interdependencies among higher-dimensional
states of neighboring agents. 

Recently, a broader category of networks termed matrix-weighted networks
has been introduced which is an immediate generalization of scalar-weighted
networks \citet{sun2018dimensional,pan2021bipartite,pan2018bipartite,trinh2018matrix,pan2020controllability,Pan2021Tac,wang2020characterizing,pan2021cluster}.
In fact, matrix-weighted networks naturally become relevant in scenarios
such as graph effective resistance based distributed control and estimation
\citet{barooah2006graph}, logical inter-dependency of multiple topics
in opinion evolution \citet{friedkin2016network}, bearing-based formation
control \citet{zhao2015translational}, array of coupled LC oscillators
\citet{tuna2016synchronization} as well as consensus and synchronization
on matrix-weighted networks \citet{trinh2018matrix,pan2018bipartite}.

As opposed to scalar-weighted networks, connectivity alone does not
translate to achieving consensus for matrix-weighted networks. To
this end, properties of weight matrices play an important role in
characterizing consensus. For instance, positive definiteness and
positive semi-definiteness of weight matrices have been employed to
provide consensus conditions in \citet{trinh2018matrix}; negative
definiteness and negative semi-definiteness of weight matrices are
further introduced in \citet{pan2018bipartite,su2019bipartite}. In
the meanwhile, the notion of network connectivity can be further extended
for matrix-weighted networks. For instance, one can identify edges
with positive/negative definite matrices as ``strong'' connections;
whereas an edge weighted by positive/negative semi-definite matrices
can be considered a ``weak'' connection \citet{trinh2018matrix,pan2018bipartite}.

During the multi-agent coordination process, simultaneous information
exchange and transmission between neighboring agents can be expensive
from the perspective of both communication and computation. The event-trigger
mechanism turns out to be efficient in handling this issue, where
the control actuation or the information transmission was determined
by the designed event \citet{ding2017overview,nowzari2019event}.
Decentralized event-triggered control for the first-order multi-agent
system was initially proposed in \citet{dimarogonas2011distributed}
which can efficiently reduce the control updates of agents. However,
the event-triggered function for the agent depends on the continuous
information monitoring of its neighbors. In order to overcome this
limitation, the distributed event-triggered functions were proposed
in \citet{nowzari2016distributed} where the state of neighboring
agents at the last event triggering time was employed to avoid the
continuous information exchange between neighboring agents. The periodically
checked event-triggered coordination strategies were addressed in
\citet{meng2013event,nowzari2016distributed}, in which the triggering
functions were only evaluated at the sampling instants avoiding the
Zeno phenomenon automatically. 

Note that the thresholds in the aforementioned results were state-dependent.
The events are triggered when the measurement error equals or exceeds
the threshold, which can be regarded as the static triggering conditions.
At the beginning of system evolution, the static triggering conditions
can effectively reduce the communication cost, as they are not easy
to be satisfied. However, as time goes by, it can be triggered frequently
since the threshold becomes smaller and smaller, leading to unnecessarily
triggered instants. In \citet{yi2018dynamic}, distributed event-triggered
consensus of the first-order multi-agent system was examined, where
a dynamic parameter associated with agents\textquoteright{} states
was introduced in the triggering conditions. The consensus of the
second-order multi-agent system was investigated in \citet{sheng2017dynamical}
with a centralized dynamic triggering condition. The distributed adaptive
dynamic event-triggered strategies for general linear multi-agent
systems are proposed in \citet{he2019adaptive}. It was shown that
dynamic parameters ensure fewer triggering instants and played essential
roles in avoiding Zeno behaviors. Recently, the event-trigger mechanism
is ubiquitously employed in distributed control of network systems.
For instance, secure synchronization of network systems using double
event-triggering mechanisms subject to actuator fault is examined
in \citet{xu2022secure}. Event-triggered security controller design
of cyber-physical systems subject to cyber-attacks are investigated
in \citet{zha2022dynamic,zha2021dynamic,liu2021quantized}. The event-triggered
dynamic output quantization controller is designed for switched T-S
fuzzy systems with unstable modes in \citet{yang2022event}. Stabilization
of markovian jump boolean networks via event-triggered control is
provided in \citet{chen2022stabilization}. The event-triggered bipartite
consensus problem for coupled general linear systems is investigated
in \citet{pan2018event,zhang2020adaptive,cai2022adaptive}. For more
details about the event-triggered problem of multi-agent systems,
one can refer to the recent survey papers and references therein \citet{ding2017overview,nowzari2019event}. 

Although the event-triggered consensus problem for scalar-weighted
networks has been extensively investigated, since matrix-weighted
networks introduce more complexity in the design and analysis of the
event-trigger mechanism, the related results are unfortunately still
in their infancy. In general, a crucial challenge in dealing with
the convergence analysis using the Lyapunov function method for matrix-weighted
networks is the loss of commutative property of edge weight compared
with scalar-weighted networks. The extension to matrix-weighted networks
is not trivial since the specific properties of weight matrices associated
with edges have to be extracted for the convenience/stability analysis
of the matrix-weighted networks under the event-triggered interaction
protocol. Recently, the event-triggered consensus problem on matrix-weighted
networks is examined in \citet{pan2021event}, where the triggering
condition is periodically checked which, however, can be computationally
inefficient. 

This paper proposes event-triggered bipartite consensus strategy for
multi-agent system on matrix-weighted networks whose edge weight allows
both positive semi-definite/definite and negative semi-definite/definite
matrices. First, a distributed event-triggered scheme with dynamic
parameters is introduced for both leaderless and leader-following
cases, the updating law of each dynamic parameter depends on the measurement
error and the relative errors between neighbors\textquoteright{} states
and its own state at triggering instants and the eigenvalue of matrix
weights between neighbors. Some sufficient conditions are derived
to guarantee leaderless and leader-following bipartite consensus,
respectively. Under these settings, the Zeno phenomenon can be avoided.
Note that the existing event-triggered consensus algorithms for scalar-weighted
networks cannot be applied to matrix-weighted networks but can be
viewed as special cases of the algorithm proposed in this paper. Finally,
simulation examples are given to verify the theoretical results.

The remainder of this paper is organized as follows. The preliminaries
of matrix analysis and graph theory are introduced in \S 2 as well
as functionality facts on consensus problem on the multi-agent system
on matrix-weighted networks. Then, the main results on the design
of event-triggered consensus protocol for leaderless matrix-weighted
networks and leader-follower matrix-weighted networks are provided
in \S 3 and \S 4, respectively, which is followed by the numerical
simulation in \S 5. The concluding remarks are finally given in \S
6.

\section{Preliminaries}

\subsection{Notations}

Let $\mathbb{R}$, $\mathbb{N}$ and $\mathbb{Z}_{+}$ be the set
of real numbers, natural numbers and positive integers, respectively.
Denote $\underline{n}=\left\{ 1,2,\ldots,n\right\} $ for an $n\in\mathbb{Z}_{+}$.
For a symmetric matrix $M$, if $M$ is positive definite (resp.,
negative definite), we write $M\succ0$ (resp., $M\prec0$); if $M$
is positive (resp., negative) semi-definite, we write $M\succeq0$
(resp., $M\preceq0$). Define the absolute value of a symmetric matrix
$M\in\mathbb{R}^{n\times n}$, denoted by $|M|$, such that $|M|=M$
if $M\succ0$ or $M\succeq0$ and $|M|=-M$ if $M\prec0$ or $M\preceq0$.
The null space of a matrix $M\in\mathbb{R}^{n\times n}$ is $\text{{\bf null}}(M)=\left\{ \boldsymbol{z}\in\mathbb{R}^{n}|M\boldsymbol{z}=\boldsymbol{0}\right\} $.
Let $\mu(M)$ denote the largest eigenvalue of a symmetric matrix
$M\in\mathbb{R}^{n\times n}$.

\subsection{Matrix-weighted Networks}

Let $\mathcal{G}=(\mathcal{V},\mathcal{E},A)$ be a matrix-weighted
network where the node set and the edge set of $\mathcal{G}$ are
denoted by $\mathcal{V}=\left\{ 1,2,\ldots,n\right\} $ and $\mathcal{E}\subseteq\mathcal{V}\times\mathcal{V}$,
respectively. The matrix-valued weight for an edge $(i,j)\in\mathcal{E}$
in $\mathcal{G}$ is a symmetric matrix $A_{ij}\in\mathbb{R}^{d\times d}$
such that $|A_{ij}|\succeq0$ or $|A_{ij}|\succ0$, and $A_{ij}=0_{d\times d}$
for all $(i,j)\not\in\mathcal{E}$. Note that a matrix-weighted network
becomes a scalar-weighted network when $d=1$. Thereby, the matrix-valued
adjacency matrix $A=[A_{ij}]\in\mathbb{R}^{dn\times dn}$ is a block
matrix such that the block located in the $i$-th row and the $j$-th
column is $A_{ij}$. We shall assume that $A_{ij}=A_{ji}$ for all
$i\not\not=j\in\mathcal{V}$ and $A_{ii}=0$ for all $i\in\mathcal{V}$,
which are analogous to the assumptions of undirected and simple graph
in a normal sense. The neighbor set of an agent $i\in\mathcal{V}$
is denoted by $\mathcal{N}_{i}=\left\{ j\in\mathcal{V}\,|\,(i,j)\in\mathcal{E}\right\} $.
Denote $D=\text{{\bf diag}}\left\{ D_{1},D_{1},\cdots,D_{n}\right\} \in\mathbb{R}^{dn}$
as the matrix-valued degree matrix of a graph where $D_{i}=\sum_{j\in\mathcal{N}_{i}}|A_{ij}|\in\mathbb{R}^{d\times d}$.
The matrix-valued Laplacian matrix of a matrix-weighted graph is defined
as $L(\mathcal{G})=D-A$, which is symmetric. A path from a node $i\in\mathcal{V}$
to a node $j\in\mathcal{V}$ in a graph $\mathcal{G}$ is a concatenation
of edges $\mathcal{P}_{i,j}=\{(i,i_{1}),(i_{1},i_{2}),\cdots,(i_{p-1},j)\}\subset\mathcal{E}$,
where all nodes $i_{1},i_{2},\ldots,i_{p}\in\mathcal{V}$ are distinct;
a node $i\in\mathcal{V}$ is reachable from a node $j\in\mathcal{V}$
if there exists a path $\mathcal{P}_{i,j}$ in $\mathcal{G}$. A graph
is connected if each pair of nodes in $\mathcal{G}$ are reachable
from each other.

Consider a multi-agent system on a matrix-weighted network $\mathcal{G}=(\mathcal{V},\mathcal{E},A)$
with $n\in\mathbb{Z}_{+}$ nodes. The state of each agent $i\in\mathcal{V}$
is denoted by $\boldsymbol{x}_{i}(t)=[x_{i1},x_{i2},\ldots,x_{id}]^{T}\in\mathbb{R}^{d}$
where $d\in\mathbb{N}$ is the size of the corresponding weight matrix.
The control protocol for each agent admits,
\begin{equation}
\dot{\boldsymbol{x}}_{i}(t)=\boldsymbol{u}_{i}(t),i\in\mathcal{V},\label{eq:the agent protocol}
\end{equation}
where
\begin{equation}
\boldsymbol{u}_{i}(t)=-\sum_{j\in\mathcal{N}_{i}}|A_{ij}|(\boldsymbol{x}_{i}(t)-\text{{\bf sgn}}(A_{ij})\boldsymbol{x}_{j}(t)),i\in\mathcal{V},
\end{equation}
the sign function $\text{{\bf sgn}}(\cdot):\mathbb{R}^{n\times n}\mapsto\left\{ 0,-1,1\right\} $
satisfies $\text{{\bf sgn}}(A_{ij})=1$ if $A_{ij}\succeq0$ or $A_{ij}\succ0$,
$\text{{\bf sgn}}(A_{ij})=-1$ if $A_{ij}\preceq0$ or $A_{ij}\prec0$,
and $\text{{\bf sgn}}(A_{ij})=0$ if $A_{ij}=0_{d\times d}$.

The overall dynamics of the multi-agent system \eqref{eq:the agent protocol}
can be characterized by the associated matrix-valued Laplacian,
\begin{equation}
\dot{\boldsymbol{x}}(t)=-L\boldsymbol{x}(t),\label{equ:matrix-consensus-overall}
\end{equation}
where $\boldsymbol{x}(t)=[\boldsymbol{x}_{1}^{T}(t),\boldsymbol{x}_{2}^{T}(t),\ldots,\boldsymbol{x}_{n}^{T}(t)]^{T}\in\mathbb{R}^{dn}$.
\begin{defn}
A bipartition of node set $\mathcal{V}$ of matrix-weighted network
$\mathcal{G}=(\mathcal{V},\mathcal{E},A)$ is two subsets of nodes
$\mathcal{V}_{i}\subset\mathcal{V}$ where $i\in\underline{2}$ such
that $\mathcal{V}=\mathcal{V}_{1}\cup\mathcal{V}_{2}$ and $\mathcal{V}_{1}\cap\mathcal{V}_{2}=\emptyset$. 
\end{defn}

In signed networks, the concept of structural balance (can be tracked
back to the seminal work\ \citet{harary1953notion}) turns out to
be an important graph-theoretic object playing a critical role in
bipartite consensus problems \citet{altafini2013consensus}. This
concept has been extended to the matrix-weighted networks in \citet{pan2018bipartite}.
\begin{defn}
\citet{pan2018bipartite}\label{def:SB-weighted-network} A matrix-weighted
network $\mathcal{G}=(\mathcal{V},\mathcal{E},A)$ is structurally
balanced if there exists a bipartition of the node set $\mathcal{V}$,
say $\mathcal{V}_{1}$ and $\mathcal{V}_{2}$, such that the matrix
weights on the edges within each subset is positive definite or positive
semi-definite, but negative definite or negative semi-definite for
the edges between the two subsets. A matrix-weighted network is structurally
imbalanced if it is not structurally balanced.
\end{defn}
Let $\mathcal{G}=(\mathcal{V},\mathcal{E},A)$ be a matrix-weighted
network with a node bipartition $\mathcal{V}_{1}$ and $\mathcal{V}_{2}$
and $d\in\mathbb{N}$ represent the dimension of edge weight. The
Gauge transformation for this node bipartition $\mathcal{V}_{1}$
and $\mathcal{V}_{2}$ is performed by the diagonal matrix $D^{*}=\text{{\bf diag}}\left\{ \sigma_{1},\sigma_{2},\ldots,\sigma_{n}\right\} $
where $\sigma_{i}=I_{d}$ if $i\in\mathcal{V}_{1}$ and $\sigma_{i}=-I_{d}$
if $i\in\mathcal{V}_{2}$. If the matrix-weighted network $\mathcal{G}=(\mathcal{V},\mathcal{E},A)$
is structurally balanced, then it satisfies that $D^{*}AD^{*}=[|A_{ij}|]\in\mathbb{R}^{dn\times dn}$. 

The following result characterizes the structure of the null space
of matrix-valued Laplacian for matrix-weighted networks, that in turn,
determine the steady-state of the multi-agent system  \eqref{equ:matrix-consensus-overall}.
\begin{lem}
\label{lem:1-1}\citet{pan2018bipartite} Let $\mathcal{G}=(\mathcal{V},\mathcal{E},A)$
be a structurally balanced matrix-weighted network. Then the Laplacian
matrix $L$ of $\mathcal{G}$ is positive semi-definite and its null
space can be characterized by
\begin{align}
\text{{\bf null}}(L) & =\text{{\bf span}}\left\{ \mathcal{R},\mathcal{H}\right\} ,\label{eq:null-space}
\end{align}
where 
\begin{equation}
\mathcal{R}=\text{{\bf range}}\{D^{*}(\boldsymbol{1}\otimes I_{d})\}\label{eq:null-space-R}
\end{equation}
and
\begin{align}
\mathcal{H=}\{\boldsymbol{v} & =[\boldsymbol{v}_{1}^{T},\boldsymbol{v}_{2}^{T},\cdots,\boldsymbol{v}_{n}^{T}]^{T}\in\mathbb{R}^{dn}\mid\nonumber \\
 & (\boldsymbol{v}_{i}-\text{{\bf sgn}}(A_{ij})\boldsymbol{v}_{j})\in\text{{\bf null}}(|A_{ij}|),\,(i,j)\in\mathcal{E}\}.\label{eq:null-space-H}
\end{align}
\end{lem}

In the following discussion, we proceed to design event-triggered
mechanisms for multi-agent system \eqref{equ:matrix-consensus-overall}
on matrix-weighted networks such that bipartite consensus can be guaranteed. 

\section{Leaderless Matrix-weighted Networks}

In this section, a distributed dynamic event-triggered coordination
strategy in the leaderless multi-agent system on the matrix-weighted
networks is discussed. First, the definition of bipartite consensus
is given.
\begin{defn}
For $i\in\mathcal{V}$ and an arbitrary $\boldsymbol{x_{i}}(0){\color{red}{\color{black}\in\mathbb{R}^{d}}}$,
the multi-agent system \eqref{equ:matrix-consensus-overall} is said
to admit bipartite consensus if ${\color{black}{\color{blue}{\color{black}\lim{}_{t\rightarrow\infty}\mid\boldsymbol{x}(t)\mid=\alpha>0}}}$.
\end{defn}
Denote by $\widehat{\boldsymbol{x}}_{i}(t)$ as the last broadcast
state of agent $i\in\mathcal{V}$ at any given time $t\geq0$, we
consider the following event-triggered control protocol, 
\begin{align}
\boldsymbol{u}_{i}(t) & =\widehat{\boldsymbol{q}}_{i}(t)\label{eq: event-trigger-protocol}\\
 & =-\sum_{j\in\mathcal{N}_{i}}|A_{ij}|(\widehat{\boldsymbol{x}}_{i}(t)-\text{{\bf sgn}}(A_{ij})\widehat{\boldsymbol{x}}_{j}(t)),\nonumber 
\end{align}
let $\widehat{\boldsymbol{x}}(t)=[\widehat{\boldsymbol{x}}_{1}^{T}(t),\widehat{\boldsymbol{x}}_{2}^{T}(t),\ldots,\widehat{\boldsymbol{x}}_{n}^{T}(t)]^{T}\in\mathbb{R}^{dn}$,
then the system \eqref{eq:the agent protocol} can be written in a
compact form as 
\begin{equation}
\dot{\boldsymbol{x}}(t)=-L\widehat{\boldsymbol{x}}(t).\label{eq:overall-event-trigger}
\end{equation}

Define the state-based measurement error between the last broadcast
state of agent $i\in\mathcal{V}$ and its current state at time $t\geq0$
as
\begin{equation}
\boldsymbol{e}_{i}(t)=\widehat{\boldsymbol{x}}_{i}(t)-\boldsymbol{x}_{i}(t),\label{eq:state measurement error-1}
\end{equation}
and the system-wise measurement error is denoted by $\boldsymbol{e}(t)=[\boldsymbol{e}_{1}^{T}(t),\boldsymbol{e}_{2}^{T}(t),\ldots,\boldsymbol{e}_{n}^{T}(t)]^{T}$. 

For agent $i\in\mathcal{V}$, the triggering time sequence is initiated
from $t_{1}^{i}=0$ and subsequently determined by

\begin{align}
t_{k+1}^{i} & =\underset{r>t_{k}^{i}}{\text{{\bf max}}}\left\{ r:\theta_{i}(\bar{\mu}_{i}\mid N_{i}\mid\parallel\boldsymbol{e}_{i}(t)\parallel^{2}-\sum_{j\in\mathcal{N}_{i}}\frac{\sigma_{i}}{4}\parallel\sqrt{|A_{ij}|}\boldsymbol{p}_{ij}\parallel^{2})\leq\chi_{i}(t),\,\forall t\in(t_{k}^{i},r]\right\} ,k\in\mathbb{Z}_{+},\label{eq:event-triggered-time}
\end{align}
where 
\begin{equation}
\boldsymbol{p}_{ij}=\widehat{\boldsymbol{x}}_{i}(t)-\text{{\bf sgn}}(A_{ij})\widehat{\boldsymbol{x}}_{j}(t),
\end{equation}
 $\bar{\mu}_{i}=\underset{j\in\mathcal{N}_{i}}{\text{{\bf max}}}\left\{ \mu(|A_{ij}|)\right\} $,
and $\sigma_{i}\in[0,1)$, $\theta_{i}$ is the parameter to be designed
and $\chi_{i}(t)$ is an auxiliary system for each agent $i\in\mathcal{V}$
such that
\begin{eqnarray}
\dot{\chi}_{i}(t) & = & -\beta_{i}\chi_{i}(t)+\delta_{i}\left(\frac{\sigma_{i}}{4}\sum_{j\in\mathcal{N}_{i}}\parallel\sqrt{|A_{ij}|}\boldsymbol{p}_{ij}\parallel^{2}-\bar{\mu}_{i}\mid N_{i}\mid\parallel\boldsymbol{e}_{i}(t)\parallel^{2}\right),\label{eq: dynamic-function}
\end{eqnarray}
with $\chi_{i}(0)>0$, $\beta_{i}>0$ and $\delta_{i}\in[0,1]$.

\begin{rem}
Note that the triggering condition \eqref{eq:event-triggered-time}
is totally distributed, it just relates to the maximal eigenvalues
of the absolute value of matrix weights between the agent and its
neighbors, it does not depend on any overall information of the system.
Also, it can be seen that only neighbors' states at triggering instants
are required which reduces the communication burden effectively.
\end{rem}
Before we proceed to present the main results in this part, the following
assumption will be employed.

\textbf{Assumption} 1\label{null-space-SB}. The matrix-weighted network
$\mathcal{G}=(\mathcal{V},\mathcal{E},A)$ is connected and structurally
balanced and there exists a Gauge transformation $D^{*}$ such that
$\text{{\bf null}}(D^{*}LD^{*})=\mathcal{R}$.
\begin{rem}
The Assumption 1 eventually guarantees that the multi-agent network
\eqref{equ:matrix-consensus-overall} admits a bipartite consensus
solution \citet{pan2018bipartite} and the bipartite solution of \eqref{equ:matrix-consensus-overall}
is 
\begin{equation}
\tilde{\boldsymbol{x}}=D^{*}(\boldsymbol{1}\otimes(\frac{1}{n}(\boldsymbol{1}^{T}\otimes I_{d})D^{*}\boldsymbol{x}(0)))
\end{equation}
 A notable distinction of matrix-weighted networks is that the network
connectivity cannot translate into achieving consensus in a multi-agent
system. In this case, even if the matrix-weighted networks is connected,
the null space of a matrix-valued Laplacian may be not equal to $\mathcal{R}$
in \ref{eq:null-space-R} (under a proper Gauge transformation), i.e.,
only the connectivity from a graph-theoretic perspective cannot guarantee
the consensus of multi-agent system in the matrix-weighted networks,
the property of the weight matrices has to be involved in the general
graph theoretic condition. Under this assumption, we shall give the
main results of this part.
\end{rem}
\begin{thm}
\label{thm:event-triggered-theorem-for-leaderless}Consider the multi-agent
system \eqref{eq:overall-event-trigger} under the matrix-weighted
network $\mathcal{G}=(\mathcal{V},\mathcal{E},A)$ satisfying \textbf{Assumption}
1. The triggering time sequence for agent $i$ is determined by \eqref{eq:event-triggered-time}.
Let $\theta_{i}$ be such that $\theta_{i}>\frac{1-\delta_{i}}{\beta_{i}}$
for all $i\in\mathcal{V}$, then the multi-agent system \eqref{eq:overall-event-trigger}
admits a bipartite consensus solution. Moreover, the Zeno behavior
can be avoided.
\end{thm}
\begin{proof}
For arbitrary $t\geq0$, according to the inequalities in \eqref{eq:event-triggered-time}
and \eqref{eq: dynamic-function}, one has 
\begin{equation}
\dot{\chi}_{i}(t)\geq-\beta_{i}\chi_{i}(t)-\frac{\delta_{i}}{\theta_{i}}\chi_{i}(t),
\end{equation}
thus, 
\begin{equation}
\chi_{i}(t)\geq\chi_{i}(0)e^{-(\beta_{i}+\frac{\delta_{i}}{\theta_{i}})t}>0.
\end{equation}
Consider the following Lyapunov function candidate $V(t)=V_{1}(t)+V_{2}(t),$
where
\begin{equation}
V_{1}(t)=\frac{1}{2}\left(\boldsymbol{x}(t)-\tilde{\boldsymbol{x}}\right)^{T}\left(\boldsymbol{x}(t)-\tilde{\boldsymbol{x}}\right),
\end{equation}
and $V_{2}(t)=\sum_{i=1}^{n}\chi_{i}(t).$ Note that 
\begin{equation}
\boldsymbol{x}(t)=-\boldsymbol{e}(t)+\widehat{\boldsymbol{x}}(t)
\end{equation}
and 
\begin{equation}
\boldsymbol{p}_{ij}=\widehat{\boldsymbol{x}}_{i}(t)-\text{{\bf sgn}}(A_{ij})\widehat{\boldsymbol{x}}_{j}(t).
\end{equation}
Then, computing the time derivative of $V_{1}(t)$ along with \eqref{eq:overall-event-trigger}
yields,

\begin{eqnarray}
\dot{V}_{1}(t) & = & \frac{1}{2}\dot{\boldsymbol{x}}^{T}(t)\boldsymbol{x}(t)+\frac{1}{2}\boldsymbol{x}(t)^{T}\dot{\boldsymbol{x}}(t)\nonumber \\
 & = & -\frac{1}{2}\widehat{\boldsymbol{x}}^{T}(t)L\boldsymbol{x}(t)-\frac{1}{2}\boldsymbol{x}(t)^{T}L\widehat{\boldsymbol{x}}(t)\nonumber \\
 & = & -\widehat{\boldsymbol{x}}(t)^{T}L\widehat{\boldsymbol{x}}(t)+\boldsymbol{e}(t)^{T}L\widehat{\boldsymbol{x}}(t)\nonumber \\
 & = & -\sum_{i=1}^{N}\sum_{j\in\mathcal{N}_{i}}\left(\frac{1}{2}\boldsymbol{p}_{ij}^{T}|A_{ij}|\boldsymbol{p}_{ij}-\boldsymbol{e}_{i}^{T}(t)|A_{ij}|\boldsymbol{p}_{ij}\right)\nonumber \\
 & \leq & -\sum_{i=1}^{N}\left(-\bar{\mu}_{i}\mid N_{i}\mid\parallel\boldsymbol{e}_{i}(t)\parallel^{2}+\sum_{j\in\mathcal{N}_{i}}\frac{1}{4}\parallel\sqrt{|A_{ij}|}\boldsymbol{p}_{ij}\parallel^{2}\right),
\end{eqnarray}
where $\bar{\mu}_{i}=\underset{j\in\mathcal{N}_{i}}{\text{{\bf max}}}\left\{ \mu(|A_{ij}|)\right\} $.

Now, we are in position to consider the Lyapunov function candidate
$V(t)$. According to the definition of $V(t)$, one has,

\begin{eqnarray}
\dot{V}(t) & = & \dot{V}_{1}(t)+\sum_{i=1}^{n}\dot{\chi}_{i}(t)\nonumber \\
 & \leq & -\sum_{i=1}^{N}\left(-\bar{\mu}_{i}\mid N_{i}\mid\parallel\boldsymbol{e}_{i}(t)\parallel^{2}+\sum_{j\in\mathcal{N}_{i}}\frac{1}{4}\parallel\sqrt{|A_{ij}|}\boldsymbol{p}_{ij}\parallel^{2}\right)\nonumber \\
 & + & \sum_{i=1}^{n}\delta_{i}\left(-\bar{\mu}_{i}\mid N_{i}\mid\parallel\boldsymbol{e}_{i}(t)\parallel^{2}+\frac{\sigma_{i}}{4}\sum_{j\in\mathcal{N}_{i}}\parallel\sqrt{|A_{ij}|}\boldsymbol{p}_{ij}\parallel^{2}\right)-\sum_{i=1}^{n}\beta_{i}\chi_{i}(t)\nonumber \\
\nonumber \\
 & \leq & \sum_{i=1}^{n}\frac{1-\delta_{i}}{\theta_{i}}\chi_{i}(t)-\sum_{i=1}^{n}\beta_{i}\chi_{i}(t)-\sum_{i=1}^{n}\frac{1-\sigma_{i}}{4}\sum_{j\in\mathcal{N}_{i}}\parallel\sqrt{|A_{ij}|}\boldsymbol{p}_{ij}\parallel^{2}\nonumber \\
 & \leq & -\sum_{i=1}^{n}\frac{1-\sigma_{max}}{4}\sum_{j\in\mathcal{N}_{i}}\parallel\sqrt{|A_{ij}|}\boldsymbol{p}_{ij}\parallel^{2}-\sum_{i=1}^{n}\left(\beta_{i}-\frac{1-\delta_{i}}{\theta_{i}}\right)\chi_{i}(t)\nonumber \\
 & \leq & -\frac{1-\sigma_{max}}{2}\widehat{\boldsymbol{x}}(t)^{T}L\widehat{\boldsymbol{x}}(t)-k_{1}\sum_{i=1}^{n}\chi_{i}(t),
\end{eqnarray}
where $\sigma_{max}=\underset{i}{\text{{\bf max}}}\left\{ \sigma_{i}\right\} $
and $k_{1}=\underset{i}{\text{{\bf min}}}\left\{ \beta_{i}-\frac{1-\delta_{i}}{\theta_{i}}\right\} $. 

Note that
\begin{eqnarray}
\boldsymbol{x}^{T}(t)L\boldsymbol{x}(t) & = & (\widehat{\boldsymbol{x}}(t)-\boldsymbol{e}(t))^{T}L(\widehat{\boldsymbol{x}}(t)-\boldsymbol{e}(t))\nonumber \\
 & \leq & 2\widehat{\boldsymbol{x}}(t)^{T}L\widehat{\boldsymbol{x}}(t)+2\parallel L\parallel\parallel\boldsymbol{e}(t)\parallel^{2}\nonumber \\
 & \leq & 2\widehat{\boldsymbol{x}}(t)^{T}L\widehat{\boldsymbol{x}}(t)+\sum_{i=1}^{n}\frac{2\parallel L\parallel\chi_{i}(t)}{\theta_{i}\bar{\mu}_{i}\mid N_{i}\mid}+\sum_{i=1}^{n}\frac{2\parallel L\parallel\sigma_{i}}{4\bar{\mu}_{i}\mid N_{i}\mid}\sum_{j\in\mathcal{N}_{i}}\parallel\sqrt{|A_{ij}|}\boldsymbol{p}_{ij}\parallel^{2}\nonumber \\
 & \leq & \left(2+\frac{\sigma_{max}\parallel L\parallel}{\underset{i}{\text{{\bf min}}}\{\bar{\mu}_{i}\mid N_{i}\mid\}}\right)\widehat{\boldsymbol{x}}(t)^{T}L\widehat{\boldsymbol{x}}(t)+\frac{2\parallel L\parallel}{\underset{i}{\text{{\bf min}}}\{\bar{\mu}_{i}\mid N_{i}\mid\theta_{i}\}}\sum_{i=1}^{n}\chi_{i}(t)\nonumber \\
 & \leq & k_{2}\widehat{\boldsymbol{x}}(t)^{T}L\widehat{\boldsymbol{x}}(t)+\frac{2\parallel L\parallel}{\underset{i}{\text{{\bf min}}}\{\bar{\mu}_{i}\mid N_{i}\mid\theta_{i}\}}\sum_{i=1}^{n}\chi_{i}(t),
\end{eqnarray}
where 
\begin{equation}
k_{2}=\text{{\bf max}}\left\{ 2+\frac{\sigma_{max}\parallel L\parallel}{\underset{i}{\text{{\bf min}}}\{\bar{\mu}_{i}\mid N_{i}\mid\}},\frac{2(1-\sigma_{max})\parallel L\parallel}{k_{1}\underset{i}{\text{{\bf min}}}\{\bar{\mu}_{i}\mid N_{i}\mid\theta_{i}\}}\right\} .
\end{equation}
 Therefore, 
\begin{eqnarray}
\widehat{\boldsymbol{x}}(t)^{T}L\widehat{\boldsymbol{x}}(t) & \geq & \frac{1}{k_{2}}\boldsymbol{x}^{T}(t)L\boldsymbol{x}(t)-\frac{2\parallel L\parallel}{k_{2}\underset{i}{\text{{\bf min}}}\{\bar{\mu}_{i}\mid N_{i}\mid\theta_{i}\}}\sum_{i=1}^{n}\chi_{i}(t),\nonumber \\
 & \geq & \frac{1}{k_{2}}\boldsymbol{x}^{T}(t)L\boldsymbol{x}(t)-\frac{k_{1}}{(1-\sigma_{max})}\sum_{i=1}^{n}\chi_{i}(t).
\end{eqnarray}
Then,
\begin{align}
-\frac{(1-\sigma_{max})}{2}\widehat{\boldsymbol{x}}(t)^{T}L\widehat{\boldsymbol{x}}(t) & \leq-\frac{(1-\sigma_{max})}{2k_{2}}\boldsymbol{x}^{T}(t)L\boldsymbol{x}(t)+\frac{k_{1}}{2}\sum_{i=1}^{n}\chi_{i}(t).
\end{align}
Thus, one further has

\begin{eqnarray}
\dot{V}(t) & \leq & -\frac{(1-\sigma_{max})}{2k_{2}}\boldsymbol{x}^{T}(t)L\boldsymbol{x}(t)-\frac{k_{1}}{2}\sum_{i=1}^{n}\chi_{i}(t)\nonumber \\
 & \leq & -\frac{\rho_{2}(L)(1-\sigma_{max})}{k_{2}}V_{1}(t)-\frac{k_{1}}{2}\sum_{i=1}^{n}\chi_{i}(t),\nonumber \\
 & \leq & -k_{3}V(t),
\end{eqnarray}
where 
\begin{equation}
k_{3}=\text{{\bf min}}\left\{ \frac{\rho_{2}(L)(1-\sigma_{max})}{k_{2}},\frac{k_{1}}{2}\right\} .
\end{equation}
Hence, 
\begin{equation}
V_{1}(t)<V(t)\leq V(0)e^{-k_{3}t},
\end{equation}
 for any $t\geq0$. Therefore, the multi-agent system \eqref{eq:overall-event-trigger}
admits a bipartite consensus solution. 

In the following, we shall prove that there is no Zeno behavior. By
contradiction, suppose that there exists Zeno behavior. Then, there
at least exists one agent $i$ such that $\text{{\bf lim}}_{k\rightarrow\infty}t_{k}^{i}=T_{0}$,
where $T_{0}>0$. From the above analysis, we know that there exists
a positive constant $M_{0}>0$ satisfying $\parallel\boldsymbol{x}_{i}(t)\parallel\leq M_{0}$
for all $t\geq0$ and $i\in\underline{n}$. Then, for any $t\geq0$,
one has 
\begin{equation}
\parallel u_{i}(t)\parallel\leq2M_{o}\sum_{j\in\mathcal{N}_{i}}\parallel A_{ij}\parallel.
\end{equation}
Choose 
\begin{equation}
\varepsilon_{0}=\left(2M_{o}\sum_{j\in\mathcal{N}_{i}}\parallel A_{ij}\parallel\right)^{-1}\sqrt{\frac{\chi_{i}(0)}{\theta_{i}\bar{\mu}_{i}\mid N_{i}\mid}}e^{-\frac{1}{2}\left(\beta_{i}+\frac{\delta_{i}}{\theta_{i}}\right)T_{0}}.
\end{equation}
Then, from the definition of limits, there exists a positive integer
$N(\varepsilon_{0})$ such that for any $k\geq N(\varepsilon_{0})$,
\begin{equation}
t_{k}^{i}\in[T_{0}-\varepsilon_{0},T_{0}].
\end{equation}
Noting that $\sum_{j\in\mathcal{N}_{i}}\frac{\sigma_{i}}{4}\parallel\sqrt{|A_{ij}|}\boldsymbol{p}_{ij}\parallel^{2}\geq0$,
we can conclude that one sufficient condition to guarantee that the
inequality in \eqref{eq:event-triggered-time} holds is

\begin{equation}
\parallel\boldsymbol{e}_{i}(t)\parallel\leq\sqrt{\frac{\chi_{i}(0)}{\theta_{i}\bar{\mu}_{i}\mid N_{i}\mid}}e^{-\frac{1}{2}(\beta_{i}+\frac{\delta_{i}}{\theta_{i}})t}.
\end{equation}
In addition, 
\begin{eqnarray}
\parallel\boldsymbol{e}_{i}(t)\parallel & = & \parallel\widehat{\boldsymbol{x}}_{i}(t_{k}^{i})-\boldsymbol{x}_{i}(t)\parallel\nonumber \\
 & = & \parallel\boldsymbol{x}_{i}(t_{k}^{i})-\boldsymbol{x}_{i}(t)\parallel\nonumber \\
 & = & \left\Vert \int_{t_{k}^{i}}^{t}\dot{\boldsymbol{x}}_{i}(t)d(t)\right\Vert \nonumber \\
 & \leq & \int_{t_{k}^{i}}^{t}\left\Vert \dot{\boldsymbol{x}}_{i}(t)\right\Vert d(t)\nonumber \\
 & \leq & 2(t-t_{k}^{i})M_{o}\sum_{j\in\mathcal{N}_{i}}\parallel A_{ij}\parallel,
\end{eqnarray}
then another sufficient condition to guarantee that the inequality
in \eqref{eq:event-triggered-time} holds is
\begin{equation}
2(t-t_{k}^{i})M_{o}\sum_{j\in\mathcal{N}_{i}}\parallel A_{ij}\parallel\leq\sqrt{\frac{\chi_{i}(0)}{\theta_{i}\bar{\mu}_{i}\mid N_{i}\mid}}e^{-\frac{1}{2}(\beta_{i}+\frac{\delta_{i}}{\theta_{i}})t}.\label{eq: second-event-triggered-time}
\end{equation}
Let $t_{N(\varepsilon_{0})+1}^{i}$ and $\tilde{t}_{N(\varepsilon_{0})+1}^{i}$
denote the next triggering time determined by the inequalities in
\eqref{eq:event-triggered-time} and \eqref{eq: second-event-triggered-time},
respectively. Then, 
\begin{eqnarray}
 &  & t_{N(\varepsilon_{0})+1}^{i}-t_{N(\varepsilon_{0})}^{i}\nonumber \\
 & \geq & \tilde{t}_{N(\varepsilon_{0})+1}^{i}-t_{N(\varepsilon_{0})}^{i}\nonumber \\
 & = & \left(2M_{o}\sum_{j\in\mathcal{N}_{i}}\parallel A_{ij}\parallel\right)^{-1}\sqrt{\frac{\chi_{i}(0)}{\theta_{i}\bar{\mu}_{i}\mid N_{i}\mid}}e^{-\frac{1}{2}\left(\beta_{i}+\frac{\delta_{i}}{\theta_{i}}\right)\tilde{t}_{N(\varepsilon_{0})+1}^{i}}\nonumber \\
 & \geq & \left(2M_{o}\sum_{j\in\mathcal{N}_{i}}\parallel A_{ij}\parallel\right)^{-1}\sqrt{\frac{\chi_{i}(0)}{\theta_{i}\bar{\mu}_{i}\mid N_{i}\mid}}e^{-\frac{1}{2}\left(\beta_{i}+\frac{\delta_{i}}{\theta_{i}}\right)T_{0}}\nonumber \\
 & = & 2\varepsilon_{0},
\end{eqnarray}
which contradicts with $t_{k}^{i}\in[T_{0}-\varepsilon_{0},T_{0}]$
for any $k\geq N(\varepsilon_{0})$. Therefore, Zero behavior is excluded.
\end{proof}
\begin{rem}
The proposed event-triggered algorithm here is not only applicable
to the matrix-weighted networks but also to the scalar-weighted networks.
Note that \eqref{eq:the agent protocol} degenerates into the scalar-weighted
case when $A_{ij}=a_{ij}I$ where $a_{ij}\in\mathbb{R}$ and $I$
denotes the $d\times d$ identity matrix and in this case, one can
choose
\begin{equation}
\bar{\mu}_{i}=\underset{j\in\mathcal{N}_{i}}{\text{{\bf max}}}\left\{ |a_{ij}|\right\} .
\end{equation}
If the scalar-valued weights associated to all edges are equal, the
triggering function \eqref{eq:event-triggered-time} is the same as
the case in the scalar-weighted networks \citet{yi2018dynamic}; otherwise,
the triggering function \eqref{eq:event-triggered-time} is easier
to be triggered than those triggering functions that are only applicable
to the scalar-weighted networks.
\end{rem}

\section{Leader-follower Matrix-weighted Networks}

Besides the leaderless network, there also exists another popular
paradigm where a subset of agents are selected as leaders or informed
agents to steer the network state to a desired one which is referred
to as leader-follower network. In a leader-follower network, a subset
of agents are referred to as leaders (or informed agents), denoted
by $\mathcal{V}_{\text{leader}}\subset\mathcal{V}$, who can be directly
influenced by the external input signal, the remaining agents are
referred to as followers, denoted by $\mathcal{V}_{\text{follower}}=\mathcal{V}\setminus\mathcal{V}_{\text{leader}}$.
The set of external input signal is denoted by $\mathcal{U}=\left\{ \boldsymbol{u}_{1},\dots,\boldsymbol{u}_{m}\right\} $
where $\boldsymbol{u}_{l}\in\mathbb{R}^{d}$, $l\in\underline{m}$
and $m\in\mathbb{Z}_{+}$. In the following discussion, we shall assume
that the input signal is homogeneous, i.e., $\boldsymbol{u}_{l_{1}}=\boldsymbol{u}_{l_{2}}=\boldsymbol{u}_{0}\ \text{for all \ensuremath{l_{1},l_{2}\in\underline{m}}}$.
Denote by the edge set between external input signals and the leaders
as $\mathcal{E}^{'}$, and a corresponding set of matrix weights as
$B=[B_{il}]\in\mathbb{R}^{nd\times md}$ where $|B_{il}|\geq0$ or
$|B_{il}|>0$ if agent $i$ is influenced by the input $\boldsymbol{u}_{l}$
and $B_{il}=0_{d\times d}$ otherwise. The graph $\bar{\mathcal{G}}=(\bar{\mathcal{V}},\bar{\mathcal{E}},\bar{A})$
is directed with $\bar{\mathcal{V}}=\mathcal{V}\cup\mathcal{U}$,
$\bar{\mathcal{E}}=\mathcal{E}\cup\mathcal{E}^{'}$, $\bar{A}=A\cup B$.
Consider the following leader-follower control protocol,
\begin{equation}
\dot{\boldsymbol{x}}_{i}(t)=\boldsymbol{u}_{i}(t),i\in\mathcal{V},\label{eq:the agent protocol-1}
\end{equation}
where
\begin{align}
\boldsymbol{u}_{i}(t) & =-\sum_{j\in\mathcal{N}_{i}}|A_{ij}|(\boldsymbol{x}_{i}(t)-\text{{\bf sgn}}(A_{ij})\boldsymbol{x}_{j}(t))\nonumber \\
 & -\sum_{l=1}^{m}|B_{il}|(\boldsymbol{x}_{i}(t)-\text{{\bf sgn}}(B_{il})\boldsymbol{u}_{l}),i\in\mathcal{V}\text{.}\label{eq:LF-unsigned-protocol}
\end{align}
The collective dynamics of \eqref{eq:the agent protocol-1} can subsequently
be characterized by 
\begin{equation}
\dot{\boldsymbol{x}}=-L_{B}(\mathcal{G})\boldsymbol{x}+B\boldsymbol{u},\label{eq: signed-LF-overall}
\end{equation}
where $\boldsymbol{x}=(\boldsymbol{x}_{1}^{T}(t),\dots,\boldsymbol{x}_{n}^{T}(t))^{T}\in\mathbb{R}^{nd}$,
$\boldsymbol{u}=(\boldsymbol{u}_{1}^{T},\dots,\boldsymbol{u}_{m}^{T})^{T}\in\mathbb{R}^{md}$
and 
\begin{equation}
L_{B}(\mathcal{G})=L(\mathcal{G})+\text{{\bf blkdiag}}(\sum_{l=1}^{m}|B_{il}|).
\end{equation}

\begin{defn}
For $i\in\mathcal{V}$ and an arbitrary $\boldsymbol{x_{i}}(0){\color{red}{\color{black}\in\mathbb{R}^{d}}}$,
the multi-agent system \eqref{eq: signed-LF-overall} is said to admit
bipartite leader-follower consensus if ${\color{black}{\color{blue}{\color{black}\lim{}_{t\rightarrow\infty}\mid\boldsymbol{x}(t)\mid=\mid\boldsymbol{u}_{0}\mid}}}$.
\end{defn}
Similar to the leaderless case, consider the following event-triggered
control protocol, 

\begin{align}
\boldsymbol{u}_{i}(t) & =\widehat{\boldsymbol{q}}_{i}(t)\label{eq:LF-event-triggered-protocol}\\
 & =-\sum_{j\in\mathcal{N}_{i}}|A_{ij}|(\widehat{\boldsymbol{x}}_{i}(t)-\text{{\bf sgn}}(A_{ij})\widehat{\boldsymbol{x}}_{j}(t))\nonumber \\
 & -\sum_{l=1}^{m}|b_{il}|(\widehat{\boldsymbol{x}}_{i}(t)-\text{{\bf sgn}}(b_{il})\boldsymbol{u}_{l}),i\in\mathcal{V},\nonumber 
\end{align}
and the system \eqref{eq:the agent protocol-1} can be written in
a compact form as 
\begin{equation}
\dot{\boldsymbol{x}}(t)=-L_{B}\widehat{\boldsymbol{x}}(t)+B\boldsymbol{u},\label{eq:LF-overall-event-trigger}
\end{equation}
where $\widehat{\boldsymbol{x}}(t)=[\widehat{\boldsymbol{x}}_{1}^{T}(t),\widehat{\boldsymbol{x}}_{2}^{T}(t),\ldots,\widehat{\boldsymbol{x}}_{n}^{T}(t)]^{T}\in\mathbb{R}^{dn}.$
Define the state-based measurement error between the last broadcast
state of agent $i\in\mathcal{V}$ and its current state at time $t\geq0$
as
\begin{equation}
\boldsymbol{e}_{i}(t)=\widehat{\boldsymbol{x}}_{i}(t)-\boldsymbol{x}_{i}(t),\label{eq:state measurement error-1-1}
\end{equation}
and the system-wise measurement error is denoted by $\boldsymbol{e}(t)=[\boldsymbol{e}_{1}^{T}(t),\boldsymbol{e}_{2}^{T}(t),\ldots,\boldsymbol{e}_{n}^{T}(t)]^{T}$. 

For agent $i\in\mathcal{V}$, the triggering time sequence is initiated
from $t_{1}^{i}=0$ and subsequently determined by

\begin{align}
t_{k+1}^{i} & =\underset{r\geq t_{k}^{i}}{\text{{\bf max}}}\left\{ r:\theta_{i}(\gamma_{i}\parallel\boldsymbol{e}_{i}(t)\parallel^{2}-\sigma_{i}\parallel\widehat{\boldsymbol{q}}_{i}(t)\parallel^{2})\leq\chi_{i}(t),\,\forall t\in[t_{k}^{i},r]\right\} ,k\in\mathbb{Z}_{+},\label{eq:event-triggered-time-1-1}
\end{align}
where $\sigma_{i}\in[0,1)$, $\theta_{i}$ and $\gamma_{i}$ are the
design parameters and $\chi_{i}(t)$ is an auxiliary system for each
agent $i\in\mathcal{V}$ such that
\begin{eqnarray}
\dot{\chi}_{i}(t) & = & -\beta_{i}\chi_{i}(t)+\delta_{i}\left(\sigma_{i}\parallel\widehat{\boldsymbol{q}}_{i}(t)\parallel^{2}-\gamma_{i}\parallel\boldsymbol{e}_{i}(t)\parallel^{2}\right),\label{eq: dynamic-function-1-1}
\end{eqnarray}
with $\chi_{i}(0)>0$, $\beta_{i}>0$ and $\delta_{i}\in[0,1]$. We
shall also denote $\sigma_{max}=\underset{i}{\text{{\bf max}}}\left\{ \sigma_{i}\right\} $
in the subsequent discussions.

\textbf{Assumption} 2.\label{null-space-SB-1} The matrix-weighted
network $\bar{\mathcal{G}}=(\bar{\mathcal{V}},\bar{\mathcal{E}},\bar{A})$
is structurally balanced and $\sum_{l=1}^{m}\sum_{i=1}^{n}|B_{il}|$
is positive definite.

The Assumption 1 and Assumption 2 together guarantee that the leader-follower
multi-agent network \eqref{eq: signed-LF-overall} admits a bipartite
leader-follower consensus. In the following, we shall give the main
results of this part.
\begin{thm}
\label{thm:event-triggered-theorem-for-leader-follower}Consider the
multi-agent system \eqref{eq:LF-overall-event-trigger} under the
matrix-weighted network $\mathcal{G}=(\mathcal{V},\mathcal{E},A)$
satisfying \textbf{Assumptions} 1 and 2. Let $\theta_{i}$ and $\gamma_{i}$
be such that $\theta_{i}>\frac{1-\delta_{i}}{\beta_{i}}$ and 
\begin{eqnarray}
\gamma_{i} & = & n\left(\sum_{j\in\mathcal{N}_{i}}\mu(\mid A_{ij}\mid)+\sum_{l=1}^{m}\mu(\mid B_{il}\mid)\right)^{2}+n\sum_{j\in\mathcal{N}_{i}}\mu^{2}(\mid A_{ij}\mid).
\end{eqnarray}
for all $i\in\mathcal{V}$, the triggering time sequence is determined
by \eqref{eq:event-triggered-time-1-1} for agent $i$ with $\chi_{i}(t)$
defined in \eqref{eq: dynamic-function-1-1}. Then the multi-agent
system \eqref{eq:LF-overall-event-trigger} admits a bipartite leader-follower
consensus. Moreover, there is no Zeno behavior.
\end{thm}
\begin{proof}
For $\forall t\geq0$, from the inequalities in \eqref{eq:event-triggered-time-1-1}
and \eqref{eq: dynamic-function-1-1}, one has 
\begin{equation}
\dot{\chi}_{i}(t)\geq-\beta_{i}\chi_{i}(t)-\frac{1}{\theta_{i}}\chi_{i}(t),
\end{equation}
thus, 
\begin{equation}
\chi_{i}(t)\geq\chi_{i}(0)e^{-(\beta_{i}+\frac{1}{\theta_{i}})t}>0.
\end{equation}
Let $\boldsymbol{\xi}(t)=\boldsymbol{x}(t)-D^{*}(\boldsymbol{1}_{n}\otimes\boldsymbol{u}_{0})$,
where $D^{*}$ is the Gauge transformation corresponding to the matrix-weighted
network $\mathcal{G}=(\mathcal{V},\mathcal{E},A)$. Then one has 

\begin{equation}
\dot{\boldsymbol{\xi}}(t)=-L_{B}\boldsymbol{\xi}(t).\label{eq:LF-error-system}
\end{equation}
Consider the Lyapunov function candidate as follows
\begin{equation}
V(t)=V_{1}(t)+V_{2}(t),
\end{equation}
where 
\begin{equation}
V_{1}(t)=\boldsymbol{\xi}^{T}(t)L_{B}\boldsymbol{\xi}(t),
\end{equation}
 and 
\begin{equation}
V_{2}(t)=\sum_{i=1}^{n}\chi_{i}(t).
\end{equation}
Computing the time derivative of $V_{1}(t)$ along with \eqref{eq:LF-error-system}
yields,
\begin{eqnarray}
\dot{V}_{1}(t) & = & \dot{\boldsymbol{\xi}}^{T}(t)L_{B}\boldsymbol{\xi}(t)+\boldsymbol{\xi}(t)^{T}L_{B}\dot{\boldsymbol{\xi}}(t)\nonumber \\
 & = & 2\boldsymbol{\xi}(t)^{T}L_{B}\widehat{\boldsymbol{q}}(t)\nonumber \\
 & = & -2\boldsymbol{q}(t)^{T}\widehat{\boldsymbol{q}}(t).
\end{eqnarray}

Define $\boldsymbol{e}_{\boldsymbol{q}(t)}(t)=\widehat{\boldsymbol{q}}(t)-\boldsymbol{q}(t)$,
then one has
\begin{eqnarray}
\dot{V}_{1}(t) & = & -2\widehat{\boldsymbol{q}}^{T}(t)\widehat{\boldsymbol{q}}(t)+2\boldsymbol{e}_{\boldsymbol{q}(t)}(t)^{T}\widehat{\boldsymbol{q}}(t)\nonumber \\
 & = & -\sum_{i=1}^{N}2\widehat{\boldsymbol{q}}_{i}^{T}(t)\widehat{\boldsymbol{q}}_{i}(t)+\sum_{i=1}^{N}2\boldsymbol{e}_{\boldsymbol{q}_{i}(t)}(t)^{T}\widehat{\boldsymbol{q}}_{i}(t)\nonumber \\
 & \leq & -\sum_{i=1}^{N}\widehat{\boldsymbol{q}}_{i}^{T}(t)\widehat{\boldsymbol{q}}_{i}(t)+\sum_{i=1}^{N}\boldsymbol{e}_{\boldsymbol{q}_{i}(t)}(t)^{T}\boldsymbol{e}_{\boldsymbol{q}_{i}(t)}(t).
\end{eqnarray}
Recal that 
\begin{eqnarray}
\boldsymbol{e}_{\boldsymbol{q}_{i}(t)}(t) & = & \sum_{j\in\mathcal{N}_{i}}|A_{ij}|\left(\text{{\bf sgn}}(A_{ij})\boldsymbol{e}_{j}(t)-\boldsymbol{e}_{i}(t)\right)-\sum_{l=1}^{m}|B_{il}|\boldsymbol{e}_{i}(t),
\end{eqnarray}
thus,

\begin{eqnarray}
\parallel\boldsymbol{e}_{\boldsymbol{q}_{i}(t)}(t)\parallel & \leq & \left(\sum_{j\in\mathcal{N}_{i}}\parallel A_{ij}\parallel+\sum_{l=1}^{m}\parallel B_{il}\parallel\right)\parallel\boldsymbol{e}_{i}(t)\parallel+\sum_{j\in\mathcal{N}_{i}}\parallel A_{ij}\parallel\parallel\boldsymbol{e}_{j}(t)\parallel,\nonumber \\
 & = & \left(\sum_{j\in\mathcal{N}_{i}}\mu(\mid A_{ij}\mid)+\sum_{l=1}^{m}\mu(\mid B_{il}\mid)\right)\parallel\boldsymbol{e}_{i}(t)\parallel+\sum_{j\in\mathcal{N}_{i}}\mu(\mid A_{ij}\mid)\parallel\boldsymbol{e}_{j}(t)\parallel.
\end{eqnarray}
Note that 
\begin{equation}
\left(\sum_{i=1}^{N}x_{i}\right)^{2}\leq N\sum_{i=1}^{N}x_{i}^{2},
\end{equation}
therefore,

\begin{eqnarray}
\parallel\boldsymbol{e}_{\boldsymbol{q}_{i}(t)}(t)\parallel^{2} & \leq & n\left(\sum_{j\in\mathcal{N}_{i}}\mu(\mid A_{ij}\mid)+\sum_{l=1}^{m}\mu(\mid B_{il}\mid)\right)^{2}\parallel\boldsymbol{e}_{i}(t)\parallel^{2}\nonumber \\
 &  & +n\sum_{j\in\mathcal{N}_{i}}\mu(\mid A_{ij}\mid)^{2}\parallel\boldsymbol{e}_{j}(t)\parallel^{2},
\end{eqnarray}
hence,
\begin{eqnarray}
\sum_{i=1}^{n}\boldsymbol{e}_{\boldsymbol{q}_{i}(t)}(t)^{T}\boldsymbol{e}_{\boldsymbol{q}_{i}(t)}(t) & \leq & \sum_{i=1}^{n}n\left(\sum_{j\in\mathcal{N}_{i}}\mu(\mid A_{ij}\mid)+\sum_{l=1}^{m}\mu(\mid B_{il}\mid)\right)^{2}\nonumber \\
 &  & \parallel\boldsymbol{e}_{i}(t)\parallel^{2}+\sum_{i=1}^{n}n\sum_{j\in\mathcal{N}_{i}}\mu^{2}(\mid A_{ij}\mid)\parallel\boldsymbol{e}_{j}(t)\parallel^{2}\nonumber \\
 & = & \sum_{i=1}^{n}\gamma_{i}\parallel\boldsymbol{e}_{i}(t)\parallel^{2}.
\end{eqnarray}
 Then, 
\begin{eqnarray}
\dot{V}_{1}(t) & \leq & \sum_{i=1}^{n}\gamma_{i}\parallel\boldsymbol{e}_{i}(t)\parallel^{2}-\sum_{i=1}^{n}\widehat{\boldsymbol{q}}_{i}^{T}(t)\widehat{\boldsymbol{q}}_{i}(t).
\end{eqnarray}
Now, we are in position to consider the Lyapunov function candidate
$V(t)$, one has

\begin{eqnarray}
\dot{V}(t) & = & \dot{V_{1}}(t)+\sum_{i=1}^{n}\dot{\chi}_{i}(t)\nonumber \\
 & \leq & \sum_{i=1}^{n}\gamma_{i}\parallel\boldsymbol{e}_{i}(t)\parallel^{2}-\sum_{i=1}^{n}\widehat{\boldsymbol{q}}_{i}^{T}(t)\widehat{\boldsymbol{q}}_{i}(t)\nonumber \\
 & + & \sum_{i=1}^{n}\left(-\beta_{i}\chi_{i}(t)+\delta_{i}(\sigma_{i}\parallel\widehat{\boldsymbol{q}}_{i}(t)\parallel^{2}-\gamma_{i}\parallel\boldsymbol{e}_{i}(t)\parallel^{2})\right)\nonumber \\
 & = & -\sum_{i=1}^{n}(1-\delta_{i}\sigma_{i})\parallel\widehat{\boldsymbol{q}}_{i}(t)\parallel^{2}+\sum_{i=1}^{n}(1-\delta_{i})\gamma_{i}\parallel\boldsymbol{e}_{i}(t)\parallel^{2}-\sum_{i=1}^{n}\beta_{i}\chi_{i}(t)\nonumber \\
 & = & \sum_{i=1}^{n}(1-\delta_{i})\gamma_{i}\parallel\boldsymbol{e}_{i}(t)\parallel^{2}-\sum_{i=1}^{n}(1-\delta_{i})\sigma_{i}\parallel\widehat{\boldsymbol{q}}_{i}(t)\parallel^{2}\nonumber \\
 & - & \sum_{i=1}^{n}\parallel\widehat{\boldsymbol{q}}_{i}(t)\parallel^{2}+\sum_{i=1}^{n}\sigma_{i}\parallel\widehat{\boldsymbol{q}}_{i}(t)\parallel^{2}-\sum_{i=1}^{n}\beta_{i}\chi_{i}(t)\nonumber \\
 & \leq & -\sum_{i=1}^{n}\left(\beta_{i}-\frac{1-\delta_{i}}{\theta_{i}}\right)\chi_{i}(t)-\sum_{i=1}^{n}(1-\sigma_{i})\parallel\widehat{\boldsymbol{q}}_{i}(t)\parallel^{2}\nonumber \\
 & \leq & -\sum_{i=1}^{n}\left(\beta_{i}-\frac{1-\delta_{i}}{\theta_{i}}\right)\chi_{i}(t)-(1-\sigma_{max})\sum_{i=1}^{n}\parallel\widehat{\boldsymbol{q}}_{i}(t)\parallel^{2}.
\end{eqnarray}
Due to $V(t)\geq0$ and $\dot{V}(t)\leq0$, which implies that $\underset{t\rightarrow\infty}{\text{{\bf lim}}}\dot{V}(t)=0$,
then one has,

\begin{eqnarray}
0 & = & \underset{t\rightarrow\infty}{\text{{\bf lim}}}\dot{V}(t)\nonumber \\
 & \leq & -\sum_{i=1}^{n}\left(\beta_{i}-\frac{1-\delta_{i}}{\theta_{i}}\right)\chi_{i}(t)-(1-\sigma_{max})\sum_{i=1}^{n}\parallel\widehat{\boldsymbol{q}}_{i}(t)\parallel^{2}\nonumber \\
 & \leq & 0,
\end{eqnarray}
thus, $\underset{t\rightarrow\infty}{\text{{\bf lim}}}\chi_{i}(t)=0$
and $\underset{t\rightarrow\infty}{\text{{\bf lim}}}\widehat{\boldsymbol{q}}_{i}(t)=\boldsymbol{0}$.
Due to 
\begin{equation}
0\leq\parallel\boldsymbol{e}_{i}(t)\parallel^{2}\leq\sigma_{i}\parallel\widehat{\boldsymbol{q}}_{i}(t)\parallel^{2}+\chi_{i}(t),
\end{equation}
therefore, $\underset{t\rightarrow\infty}{\text{{\bf lim}}}\boldsymbol{e}_{i}(t)=\boldsymbol{0}$.
Then, 
\begin{eqnarray}
\dot{V}_{1}(t) & = & \dot{\boldsymbol{\xi}}^{T}(t)L_{B}\boldsymbol{\xi}(t)+\boldsymbol{\xi}(t)^{T}L_{B}\dot{\boldsymbol{\xi}}(t)\nonumber \\
 & = & -2\boldsymbol{\xi}(t)^{T}L_{B}L_{B}(\boldsymbol{\xi}(t)+\boldsymbol{e}(t))\nonumber \\
 & = & -\boldsymbol{\xi}(t)^{T}L_{B}L_{B}\boldsymbol{\xi}(t)-\boldsymbol{\xi}(t)^{T}L_{B}L_{B}\boldsymbol{e}(t),
\end{eqnarray}
thus, $\underset{t\rightarrow\infty}{\text{{\bf lim}}}L_{B}\boldsymbol{\xi}(t)=\boldsymbol{0}$,
when the interaction graph is connected, the matrix $L_{B}$ is positive
definite, therefore,
\begin{equation}
\underset{t\rightarrow\infty}{\text{{\bf lim}}}\boldsymbol{\xi}(t)=\boldsymbol{0},
\end{equation}
then the multi-agent system \eqref{eq:LF-overall-event-trigger} admits
a bipartite leader-follower consensus.

In the following discussion, we shall prove that there is no Zeno
behavior. By contradiction, suppose that there exists Zeno behavior
when applying the proposed event-trigger control protocol \eqref{eq:LF-event-triggered-protocol}
to the multi-agent system. Then, there at least exists one agent $i$
such that $\text{{\bf lim}}_{k\rightarrow\infty}t_{k}^{i}=T_{0}$
where $T_{0}>0$. From the above analysis, we know that there exists
a positive constant $M_{0}>0$ satisfying $\parallel\boldsymbol{x}_{i}(t)\parallel\leq M_{0}$
for all $t\geq0$ and $i\in\underline{n}$. Then one has 
\begin{equation}
\parallel\boldsymbol{p}_{i}(t)\parallel\leq2M_{o}\sum_{j\in\mathcal{N}_{i}}\parallel A_{ij}\parallel+(M_{0}+\parallel\boldsymbol{u}_{0}\parallel)\sum_{l=1}^{m}\parallel B_{il}\parallel,
\end{equation}
 for any $t\geq0$. Choose 
\begin{align}
\varepsilon_{0} & =\left(2M_{o}\sum_{j\in\mathcal{N}_{i}}\parallel A_{ij}\parallel(M_{0}+\parallel\boldsymbol{u}_{0}\parallel)\sum_{l=1}^{m}\parallel B_{il}\parallel\right)^{-1}\sqrt{\frac{\chi_{i}(0)}{\theta_{i}r_{i}}}e^{-\frac{1}{2}(\beta_{i}+\frac{\delta_{i}}{\theta_{i}})T_{0}}.
\end{align}
Then, according to the definition of limits, there exists a positive
integer $N(\varepsilon_{0})$ such that for any $k\geq N(\varepsilon_{0})$,
\begin{equation}
t_{k}^{i}\in[T_{0}-\varepsilon_{0},T_{0}].\label{eq:the limit}
\end{equation}
Note that $\sigma_{i}\parallel\boldsymbol{p}_{i}(t)\parallel^{2}\geq0$,
then one sufficient condition to guarantee the inequality in \eqref{eq:event-triggered-time-1-1}
is

\begin{equation}
\parallel\boldsymbol{e}_{i}(t)\parallel\leq\sqrt{\frac{\chi_{i}(0)}{\theta_{i}r_{i}}}e^{-\frac{1}{2}(\beta_{i}+\frac{1}{\theta_{i}})t}.
\end{equation}
In addition, 
\begin{eqnarray}
\parallel\boldsymbol{e}_{i}(t)\parallel & = & \parallel\widehat{\boldsymbol{x}}_{i}(t_{k}^{i})-\boldsymbol{x}_{i}(t)\parallel\nonumber \\
 & = & \parallel\boldsymbol{x}_{i}(t_{k}^{i})-\boldsymbol{x}_{i}(t)\parallel\nonumber \\
 & = & \left\Vert \int_{t_{k}^{i}}^{t}\dot{\boldsymbol{x}}_{i}(t)d(t)\right\Vert \nonumber \\
 & \leq & \int_{t_{k}^{i}}^{t}\parallel\dot{\boldsymbol{x}}_{i}(t)\parallel d(t)\nonumber \\
 & \leq & (t-t_{k}^{i})(2M_{o}\sum_{j\in\mathcal{N}_{i}}\parallel A_{ij}\parallel+(M_{0}+\parallel\boldsymbol{u}_{0}\parallel)\sum_{l=1}^{m}\parallel b_{il}\parallel),
\end{eqnarray}
then another sufficient condition to guarantee that the inequality
in \eqref{eq:event-triggered-time-1-1} holds if
\begin{align}
(t-t_{k}^{i})(2M_{o}\sum_{j\in\mathcal{N}_{i}} & \parallel A_{ij}\parallel\nonumber \\
+ & (M_{0}+\parallel\boldsymbol{u}_{0}\parallel)\sum_{l=1}^{m}\parallel B_{il}\parallel)\nonumber \\
\leq & \sqrt{\frac{\chi_{i}(0)}{\theta_{i}r_{i}}}e^{-\frac{1}{2}(\beta_{i}+\frac{1}{\theta_{i}})t}.\label{eq:second-event-triggered-time-1}
\end{align}
Let $t_{N(\varepsilon_{0})+1}^{i}$ and $\tilde{t}_{N(\varepsilon_{0})+1}^{i}$
denote the next triggering time determined by the inequalities in
\eqref{eq:event-triggered-time-1-1} and \eqref{eq:second-event-triggered-time-1},
respectively. Then, 

\begin{eqnarray}
t_{N(\varepsilon_{0})+1}^{i}-t_{N(\varepsilon_{0})}^{i} & \geq & \tilde{t}_{N(\varepsilon_{0})+1}^{i}-t_{N(\varepsilon_{0})}^{i}\nonumber \\
 & = & \left(2M_{o}\sum_{j\in\mathcal{N}_{i}}\parallel A_{ij}\parallel(M_{0}+\parallel\boldsymbol{u}_{0}\parallel)\sum_{l=1}^{m}\parallel B_{il}\parallel\right)^{-1}\sqrt{\frac{\chi_{i}(0)}{\theta_{i}r_{i}}}e^{-\frac{1}{2}(\beta_{i}+\frac{1}{\theta_{i}})\tilde{t}_{N(\varepsilon_{0})+1}^{i}}\nonumber \\
 & \geq & \left(2M_{o}\sum_{j\in\mathcal{N}_{i}}\parallel A_{ij}\parallel+(M_{0}+\parallel\boldsymbol{u}_{0}\parallel)\sum_{l=1}^{m}\parallel B_{il}\parallel\right)^{-1}\sqrt{\frac{\chi_{i}(0)}{\theta_{i}r_{i}}}e^{-\frac{1}{2}(\beta_{i}+\frac{\delta_{i}}{\theta_{i}})T_{0}}\nonumber \\
 & = & 2\varepsilon_{0},
\end{eqnarray}
which contradicts with the equation in \eqref{eq:the limit}. Therefore,
Zeno behavior is excluded.
\end{proof}
\begin{rem}
Similar to the leaderless case, the multi-agent system \eqref{eq: signed-LF-overall}
degenerates into the scalar-weighted case when $A_{ij}=a_{ij}I$,
in this case, 
\begin{equation}
\gamma_{i}=n\left(\sum_{j\in\mathcal{N}_{i}}\mid a_{ij}\mid+\sum_{l=1}^{m}\mid b_{il}\mid\right)^{2}+n\sum_{j\in\mathcal{N}_{i}}a_{ij}^{2}.
\end{equation}
Therefore, the event-triggered strategy proposed for the matrix-weighted
leader-follower system can be applied for the scalar-weighted leader-follower
case directly.
\end{rem}

\section{Simulations}

In this section, we proceed to provide two simulation examples to
demonstrate the theoretical results in this paper. 

\subsection{Leaderless Matrix-weighted Networks}

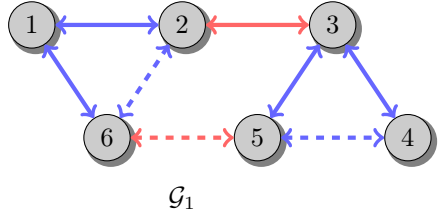
\begin{figure}
\begin{centering}
\begin{tikzpicture}[scale=1]

	\node (n1) at (0,1.5) [circle,circular drop shadow,fill=black!20,draw] {1};
    \node (n2) at (2,1.5) [circle,circular drop shadow,fill=black!20,draw] {2};
    \node (n3) at (4,1.5) [circle,circular drop shadow,fill=black!20,draw] {3};
	\node (n4) at (5,0) [circle,circular drop shadow,fill=black!20,draw] {4};
    \node (n5) at (3,0) [circle,circular drop shadow,fill=black!20,draw] {5};
    \node (n6) at (1,0) [circle,circular drop shadow,fill=black!20,draw] {6};


	\node (G) at (2,-0.8) {$\mathcal{G}_1$};


	\draw[<->, ultra thick, color=red!60] (n2) -- (n3); 
	\draw[<->, ultra thick, color=red!60,dashed] (n6) -- (n5);

	\draw [<->, ultra thick, color=blue!60] (n1) -- (n2); 
	\draw [<->, ultra thick, color=blue!60] (n1) -- (n6); 
	\draw [<->, ultra thick, color=blue!60,dashed] (n2) -- (n6); 
	\draw [<->, ultra thick, color=blue!60] (n5) -- (n3); 
	\draw [<->, ultra thick, color=blue!60] (n3) -- (n4); 
	\draw [<->, ultra thick, color=blue!60,dashed] (n5) -- (n4); 
\end{tikzpicture}
\par\end{centering}
\caption{A six-node structurally balanced matrix-weighted network $\mathcal{G}_{1}$.
The solid lines represent the edges weighted by (positive or negative)
definite matrices, the dashed lines represent the edges weighted by
(positive or negative) semi-definite matrices. The blue lines represent
edges weighted by positive (semi-)definite matrices, and red lines
represent edges weighted by negative (semi-)definite matrices.}

\label{fig:Figure1-1}
\end{figure}

First, consider the leaderless multi-agent system \eqref{eq:overall-event-trigger}
on the structurally balanced matrix-weighted network $\mathcal{G}_{1}$
with the node bipartition $\mathcal{V}_{1}=\{1,2,6\}$ and $\mathcal{V}_{2}=\{3,4,5\}$
as shown in Figure \ref{fig:Figure1-1}. 

\begin{figure}[h]
\begin{centering}
\includegraphics[width=12cm]{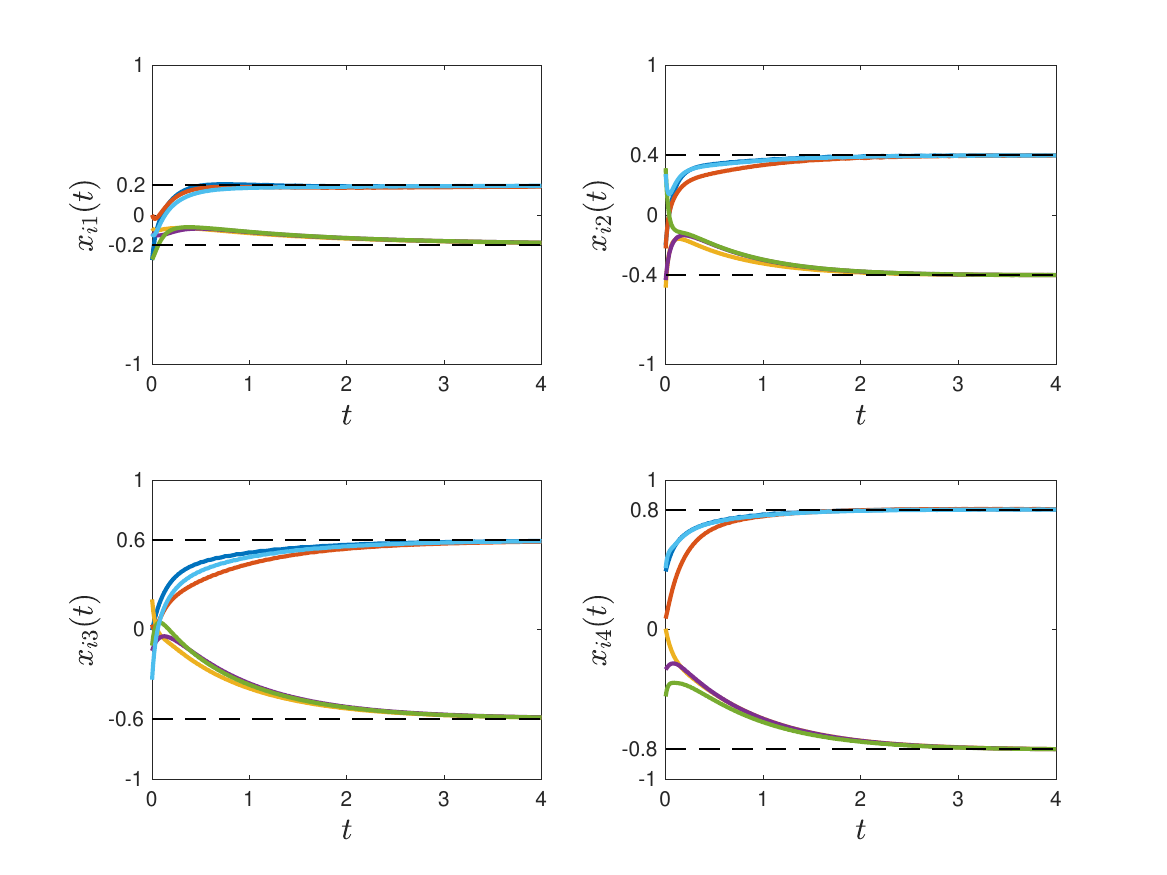}
\par\end{centering}
\caption{Entry-wise trajectory of each agent for the multi-agent system \eqref{eq:overall-event-trigger}
under the structurally balanced matrix-weighted network $\mathcal{G}_{1}$
in Figure \ref{fig:Figure1-1}.}

\label{fig:LL-trajectory}
\end{figure}

In this case, the state dimension of each agent is $d=4$, and all
agents adopt event-triggered control protocol \eqref{eq: event-trigger-protocol}.
The weight matrices on edges in $\mathcal{G}_{1}$ are
\begin{figure}[H]
\begin{centering}
\includegraphics[width=12cm]{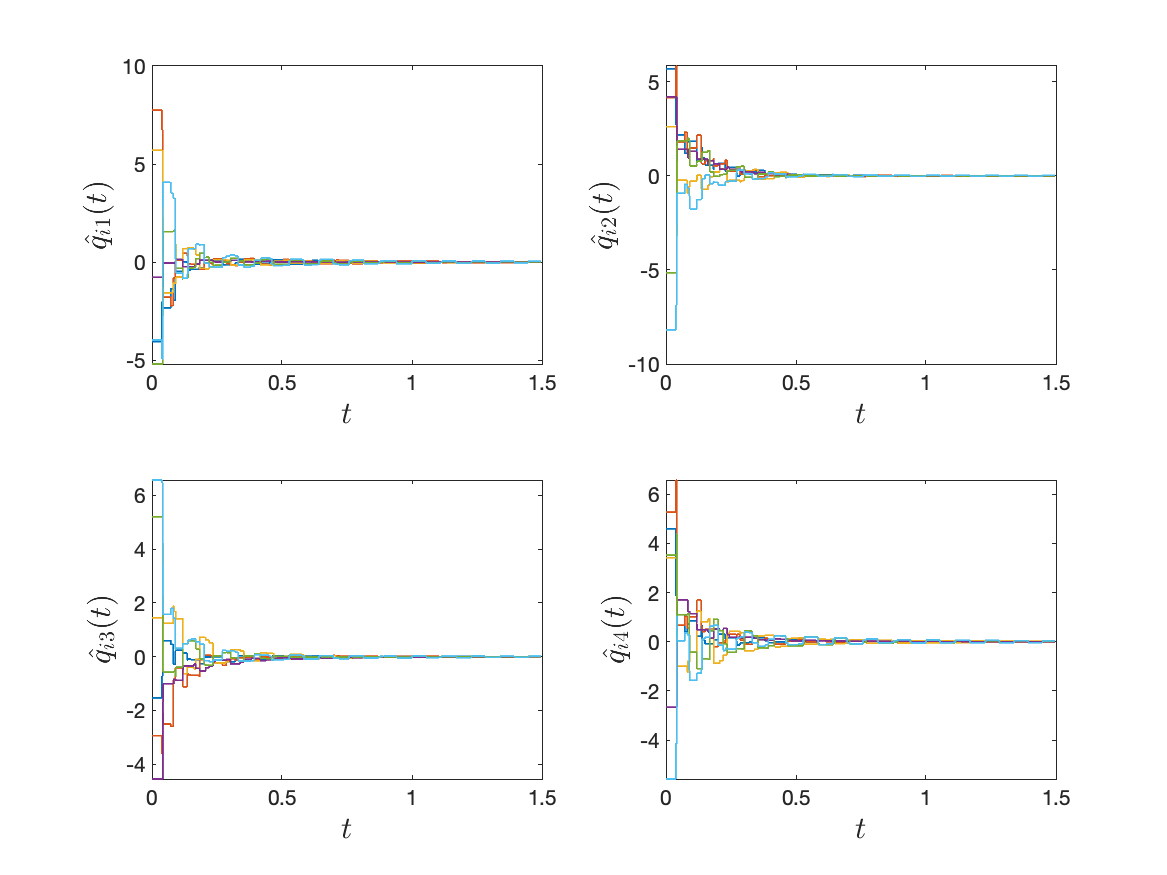}
\par\end{centering}
\caption{The event-based control protocol $\widehat{\boldsymbol{q}}_{i}(t)$
of each agent $i\in\mathcal{V}$ for the multi-agent system \eqref{eq:overall-event-trigger}
under the structurally balanced matrix-weighted network $\mathcal{G}_{1}$.}

\label{fig:LL-trigger-control}
\end{figure}

\[
A_{12}=\left[\begin{array}{cccc}
0.0975 & 0.9649 & 0.4854 & 0.9157\\
0.2785 & 0.1576 & 0.8003 & 0.7922\\
0.5469 & 0.9706 & 0.1419 & 0.9595\\
0.9575 & 0.9572 & 0.4218 & 0.6557
\end{array}\right]>0,
\]

\[
A_{16}=\begin{bmatrix}8.1684 & 1 & -0.1160 & 0.3328\\
1 & 6.7495 & 1.2264 & 0.4473\\
-0.1160 & 1.2264 & 7.4303 & 0.2236\\
0.3328 & 0.4473 & 0.2236 & 8.0775
\end{bmatrix}>0,
\]

\[
A_{26}=\begin{bmatrix}4.6211 & 0.8971 & 0.8392 & 2.7045\\
0.8971 & 1.1161 & 2.1934 & 0.0274\\
0.8392 & 2.1934 & 4.5295 & -0.5815\\
2.7045 & 0.0274 & -0.5815 & 1.8457
\end{bmatrix}\ge0,
\]

\[
A_{23}=\begin{bmatrix}-6.6469 & 0.4166 & 0.044 & 0.2922\\
0.4166 & -8.2131 & 0.1152 & -0.3055\\
0.044 & 0.1152 & -6.2339 & -0.1434\\
0.2922 & -0.3055 & -0.1434 & -6.6147
\end{bmatrix}<0,
\]

\[
A_{56}=\begin{bmatrix}-4.7176 & -1.6485 & 1.5246 & -3.1114\\
-1.6485 & -6.7837 & -1.3214 & 0.9421\\
1.5246 & -1.3214 & -6.4716 & -2.6201\\
-3.1114 & 0.9421 & -2.6201 & -6.0166
\end{bmatrix}\le0,
\]

\[
A_{35}=\begin{bmatrix}4.8630 & -0.9583 & -1.0002 & 0.6242\\
-0.9583 & 4.9516 & 1.1961 & -0.8268\\
-1.0002 & 1.1961 & 6.5071 & -2.4257\\
0.6242 & -0.8268 & -2.4257 & 6.4197
\end{bmatrix}>0,
\]
\[
A_{34}=\begin{bmatrix}4.6843 & -0.5024 & 1.2292 & 0.5247\\
-0.5024 & 6.2876 & 0.5766 & 0.0968\\
1.2292 & 0.5766 & 5.2446 & 0.0118\\
0.5247 & 0.0968 & 0.0118 & 6.2167
\end{bmatrix}>0,
\]
and

\[
A_{45}=\begin{bmatrix}0.7899 & 1.5860 & -0.3137 & -0.498\\
1.5860 & 3.2857 & -1.0541 & -1.5607\\
-0.3137 & -1.0541 & 1.9019 & 2.5477\\
-0.4980 & -1.5607 & 2.5477 & 3.4211
\end{bmatrix}\ge0.
\]
Moreover, $A_{ij}=A_{ji}$ for all $(i,j)\in\mathcal{E}(\mathcal{G}_{1})$.
Choose $\sigma_{i}=0.9$, $\delta_{i}=1$, $\beta_{i}=1$, and $\chi_{i}(0)=0.5$.
According to \textbf{Theorem} \ref{thm:event-triggered-theorem-for-leaderless},
choose $\theta_{i}=0.5$ which satisfies $\theta_{i}>\frac{1-\delta_{i}}{\beta_{i}}$.
Each dimension of initial value corresponding to each agent is randomly
chosen from $[-1,1]$. By computing the eigenvalues of the weight
matrices, one can get $\bar{\mu}_{1}=9.2047,\,\bar{\mu}_{2}=8.396,\,\bar{\mu}_{3}=9.7599,\,\bar{\mu}_{4}=6.7454,\,\bar{\mu}_{5}=9.7599,\,\bar{\mu}_{6}=9.3996$.
Using the above parameters, the bipartite consensus can be achieved
in an element-wise manner, as shown in Figure \ref{fig:LL-trajectory}.
The dimensions of control protocol for each agent are illustrated
in Figure \ref{fig:LL-trigger-control}.  Sequences of triggering
time for each agent are illustrated in Figure \ref{fig:LL-trigger-time}.

\begin{figure}[H]
\begin{centering}
\includegraphics[width=12cm]{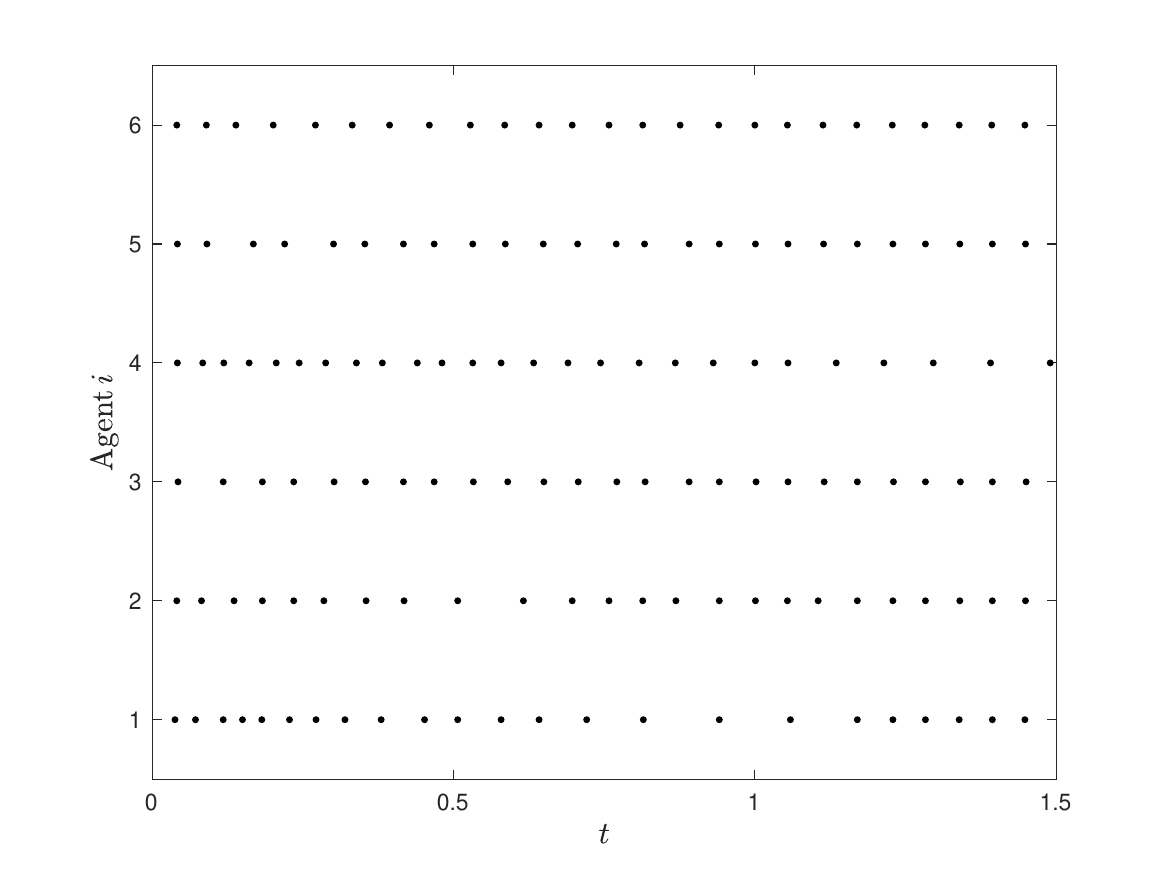}
\par\end{centering}
\caption{The triggering time instants of each agent in the multi-agent system
\eqref{eq:overall-event-trigger} under the structurally balanced
matrix-weighted network $\mathcal{G}_{1}$.}

\label{fig:LL-trigger-time}
\end{figure}

\subsection{Leader-follower Matrix-weighted Networks}

\begin{figure}
\begin{centering}
\begin{tikzpicture}[scale=1]

	\node (n1) at (0,1.5) [circle,circular drop shadow,fill=black!20,draw] {1};
    \node (n2) at (2,1.5) [circle,circular drop shadow,fill=black!20,draw] {2};
    \node (n3) at (4,1.5) [circle,circular drop shadow,fill=black!20,draw] {3};
	\node (n4) at (5,0) [circle,circular drop shadow,fill=black!20,draw] {4};
    \node (n5) at (3,0) [circle,circular drop shadow,fill=black!20,draw] {5};
    \node (n6) at (1,0) [circle,circular drop shadow,fill=black!20,draw] {6};

    \node (u1) at (-1.7,1.6) [circle,inner sep= 1.5pt,fill=black!20,draw] {$u_1$};
    \node (u2) at (-0.7,0.1) [circle,inner sep= 1.5pt,fill=black!20,draw] {$u_2$};

	\draw[->, ultra thick, dashed, color=blue!60]  (u1) -- (n1); 
	\draw[->, ultra thick, color=blue!60]  (u2) -- (n6); 

	\node (G) at (2,-0.8) {$\mathcal{G}^{\prime}_1$};


	\draw[<->, ultra thick, color=red!60] (n2) -- (n3); 
	\draw[<->, ultra thick, color=red!60,dashed] (n6) -- (n5);

	\draw [<->, ultra thick, color=blue!60] (n1) -- (n2); 
	\draw [<->, ultra thick, color=blue!60] (n1) -- (n6); 
	\draw [<->, ultra thick, color=blue!60,dashed] (n2) -- (n6); 
	\draw [<->, ultra thick, color=blue!60] (n5) -- (n3); 
	\draw [<->, ultra thick, color=blue!60] (n3) -- (n4); 
	\draw [<->, ultra thick, color=blue!60,dashed] (n5) -- (n4); 
\end{tikzpicture}
\par\end{centering}
\caption{A six-node structurally balanced matrix-weighted network with two
external inputs $\boldsymbol{u}_{1}$ and $\boldsymbol{u}_{2}$, denoted
by $\mathcal{G}_{1}^{\prime}$. The correspondence between the line
pattern and weight matrix is the same as that in Figure \ref{fig:Figure1-1}.}

\label{fig:LF-network}
\end{figure}
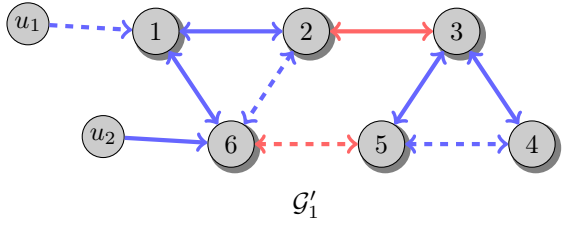

Consider the leader-follower multi-agent system \eqref{eq:LF-overall-event-trigger}
on the leader-follower network $\mathcal{G}_{1}^{\prime}$ in Figure
\ref{fig:LF-network}, where agents $1$ and $6$ are the leaders
influenced by the inputs $\boldsymbol{u}_{1}$ and $\boldsymbol{u}_{2}$,
respectively. The edge weights in the matrix-weighted network $\mathcal{G}_{1}^{\prime}$
are the same as the leaderless case above, the influence weights by
the inputs $\boldsymbol{u}_{1}$ and $\boldsymbol{u}_{2}$ are $B_{11}=A_{45}\ge0$
and $B_{62}=A_{16}>0$, respectively. 
 In this case, choose $\sigma_{i}=0.9$, $\delta_{i}=1$, $\beta_{i}=1$,
$\chi_{i}(0)=0.5$, and $\boldsymbol{u}_{1}=\boldsymbol{u}_{2}=[0.2,0.4,0.6,0.8]^{T}$.
By computing the eigenvalues of the weight matrices, one has $\gamma_{1}=9.2047,\,\gamma_{2}=8.396,\,\gamma_{3}=9.7599,\,\gamma_{4}=6.7454,\,\gamma_{5}=9.7599,\,\gamma_{6}=9.3996$.
\begin{figure}[h]
\begin{centering}
\includegraphics[width=12cm]{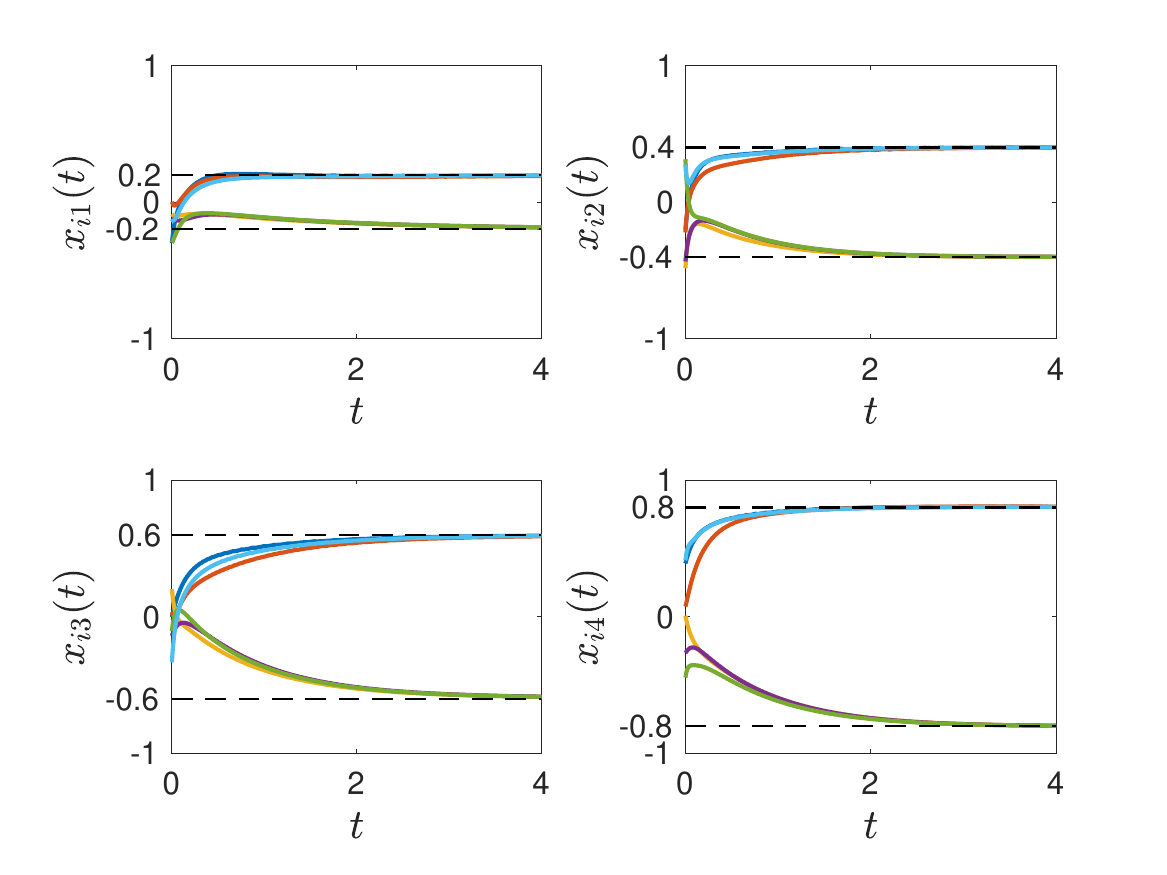}
\par\end{centering}
\caption{Entry-wise trajectory of each agent for the multi-agent system \eqref{eq:LF-overall-event-trigger}
under the leader-follower network $\mathcal{G}_{1}^{\prime}$ in Figure
\ref{fig:LF-network}.}

\label{fig:LF-trajectory}
\end{figure}
 According to \textbf{Theorem} \ref{thm:event-triggered-theorem-for-leader-follower},
choose $\theta_{i}=1$ satisfying $\theta_{i}>\frac{1-\delta_{i}}{\beta_{i}}$.
Under these parameters, the bipartite leader-follower consensus can
be achieved as shown in Figure \ref{fig:LF-trajectory}. 
\begin{figure}[H]
\begin{centering}
\includegraphics[width=12cm]{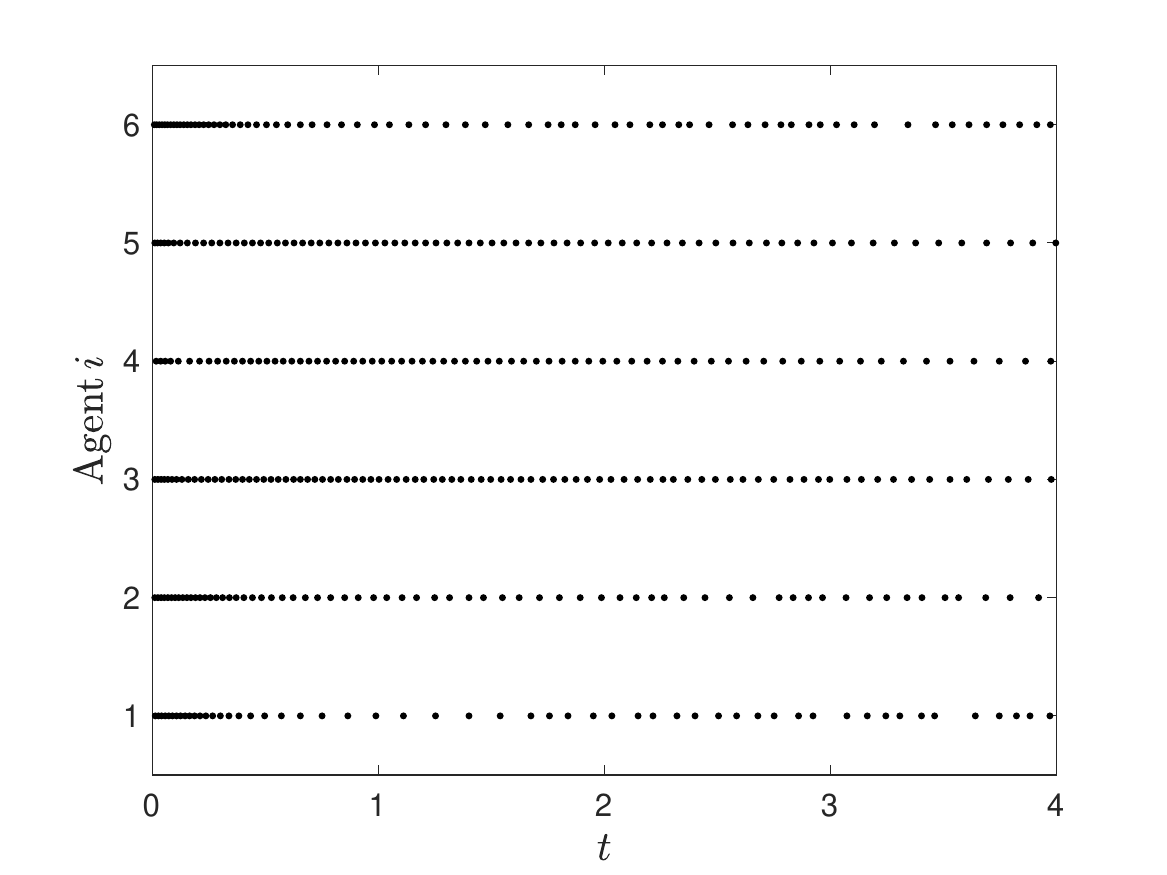}
\par\end{centering}
\caption{The triggering time instants of each agent in the multi-agent system
\eqref{eq:LF-overall-event-trigger} under the leader-follower network
$\mathcal{G}_{1}^{\prime}$ in Figure \ref{fig:LF-network}.}

\label{fig:LF-trigger-time}
\end{figure}
Sequences of triggering time for each agent are demonstrated in Figure
\ref{fig:LF-trigger-time}. The dimensions of control protocol for
each agent are illustrated in Figure \ref{fig:LF-trigger-control}.

\begin{figure}[H]
\begin{centering}
\includegraphics[width=12cm]{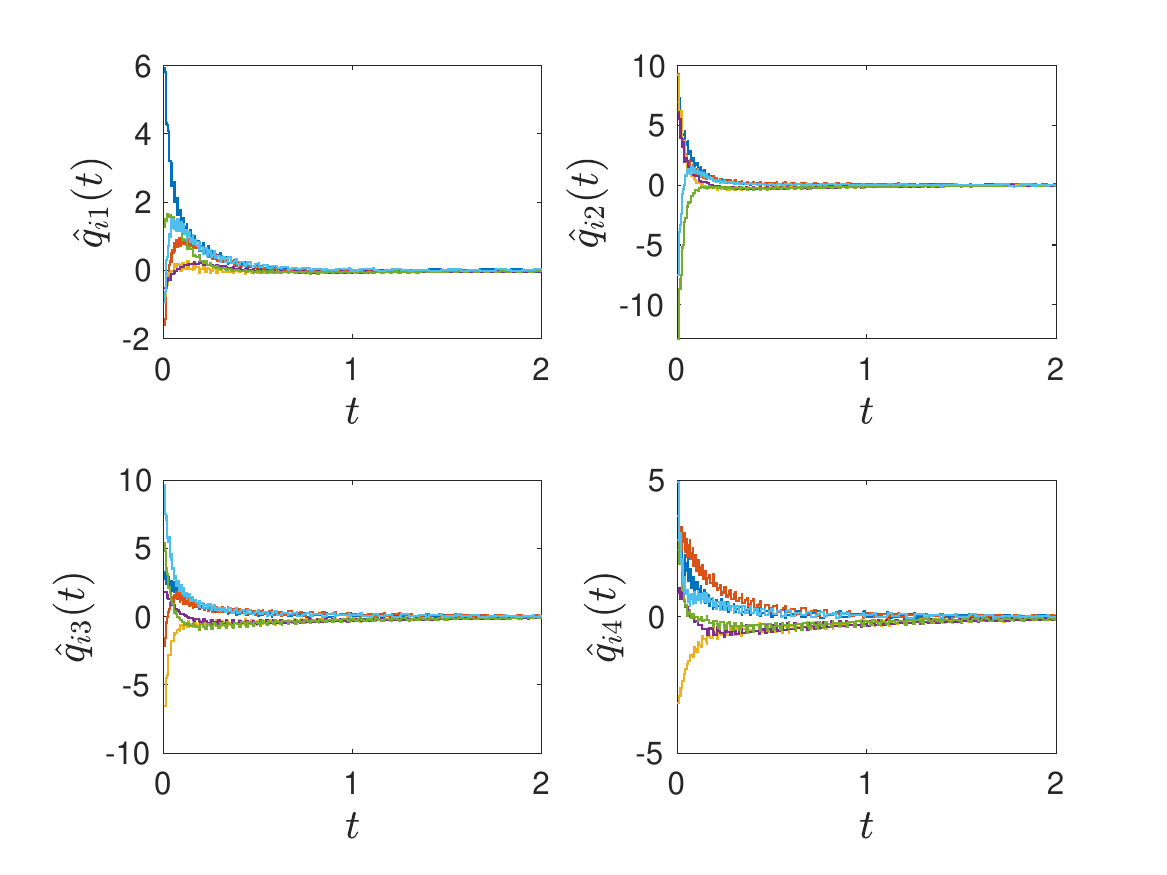}
\par\end{centering}
\caption{The event-based control protocol\textcolor{black}{{} $\widehat{\boldsymbol{q}}_{i}(t)$
}of each agent $i\in\mathcal{V}$ for the multi-agent system \eqref{eq:LF-overall-event-trigger}
under the leader-follower network $\mathcal{G}_{1}^{\prime}$ in Figure
\ref{fig:LF-network}.}

\label{fig:LF-trigger-control}
\end{figure}

\section{Conclusion }

The event-triggered bipartite consensus strategies for both leaderless
and leader-follower multi-agent system on matrix-weighted networks
are discussed in this paper. By introducing an additional variable,
which is generated by an auxiliary system, for each agent to adjust
its threshold dynamically, the proposed distributed dynamic event-triggered
strategies are proposed for the matrix-weighted networks and the Zeno
behavior for the triggering time sequence can be avoided. In the proposed
event-triggered strategies, each agent only needs to broadcast at
its own triggering instants, and listen to incoming information from
its neighbors at their triggering instants. Thus, continuous measurement
of neighbors' states can be avoided. Simulation examples are provided
to demonstrate the theoretical results. The future work includes the
extension of the proposed algorithm to directed networks and event-triggered
consensus problem of high-order matrix-weighted networks.

\section{Appendix}
\begin{lem}
\label{lem:1}Let $\boldsymbol{x},\,\boldsymbol{y}\in\mathbb{R}^{d}$
and $\alpha>0$. Then 
\[
\boldsymbol{x}^{T}\boldsymbol{y}\leq\frac{\boldsymbol{x}^{T}\boldsymbol{x}}{2\alpha}+\frac{\alpha\boldsymbol{y}^{T}\boldsymbol{y}}{2}.
\]
\end{lem}
\begin{lem}
\label{lem:Rayleigh Theorem}\citet{horn2012matrix} Let $M\in\mathbb{R}^{n\times n}$
be Hermitian with eigenvalues $\lambda_{1}\leq\cdots\leq\lambda_{n}$.
Let $\boldsymbol{x}_{i_{1}},\cdots,\boldsymbol{x}_{i_{k}}$ be mutually
orthonormal vectors such that $M\boldsymbol{x}_{i_{p}}=\lambda_{i_{p}}\boldsymbol{x}_{i_{p}}$
where $i_{p}\in\mathbb{N}$, $p\in\underline{k}$ and $1\leq i_{1}<\cdots<i_{k}\leq n$.
Then
\[
\lambda_{i_{1}}=\underset{\{\boldsymbol{x}\not={\bf 0},\boldsymbol{x}\in S\}}{\text{{\bf min}}}\frac{\boldsymbol{x}^{\top}M\boldsymbol{x}}{\boldsymbol{x}^{\top}\boldsymbol{x}},
\]
and
\[
\lambda_{i_{k}}=\underset{\{\boldsymbol{x}\not={\bf 0},\boldsymbol{x}\in S\}}{\text{{\bf max}}}\frac{\boldsymbol{x}^{\top}M\boldsymbol{x}}{\boldsymbol{x}^{\top}\boldsymbol{x}},
\]
where $S=\text{{\bf span}}\{\boldsymbol{x}_{i_{1}},\cdots,\boldsymbol{x}_{i_{k}}\}$.
\end{lem}
\bibliographystyle{elsarticle-harv}
\addcontentsline{toc}{section}{\refname}\bibliography{mybib,mybib-dynamic-event-trigger}

\end{document}